\documentclass[aps,prl,superscriptaddress,notitlepage,twocolumn]{revtex4-1}
\usepackage[colorlinks=true,bookmarks=false,citecolor=blue,urlcolor=blue]{hyperref}   
\usepackage{amsmath}
\usepackage{graphicx}
\usepackage{color}
\usepackage{braket}

\newcommand{\be}{\begin{equation}}
\newcommand{\ee}{\end{equation}}

\usepackage{bm}

\newcommand{\bs}{\boldsymbol}

\newcommand{\revision}{}
\newcommand{\revisiontwo}{}

\usepackage{upgreek}
   
\usepackage{verbatim} 
\usepackage{morefloats}
\usepackage[mathscr]{euscript}
\usepackage{enumerate}
\usepackage{amsfonts}

\renewcommand{\Im}{\mathrm{Im}}
\def\dfrac{\displaystyle\frac}  

\newcommand{\pcsadd}{Center for Theoretical Physics of Complex Systems, Institute for Basic Science, Daejeon 34126, Korea}
\newcommand{\ustadd}{Basic Science Program, Korea University of Science and Technology, Daejeon 34113, Korea}
\newcommand{\vincaadd}{P$^{*}$ Group, Vin\v{c}a Institute of Nuclear Sciences, University of Belgrade, P.O. Box 522, 11001 Belgrade, Serbia}
\begin{document}

\title{Probing band topology using modulational instability}

\author{Daniel Leykam}
\altaffiliation{Present address: Centre for Quantum Technologies, National University of Singapore, 3 Science Drive 2, Singapore 117543}
\affiliation{\pcsadd}
\affiliation{\ustadd}

\author{Ekaterina Smolina}
\affiliation{Institute of Applied Physics, Russian Academy of Science, Nizhny Novgorod 603950, Russia}

\author{Aleksandra Maluckov}
\affiliation{\pcsadd}
\affiliation{\vincaadd}

\author{Sergej Flach}
\affiliation{\pcsadd}
\affiliation{\ustadd}

\author{Daria A. Smirnova}
\affiliation{Institute of Applied Physics, Russian Academy of Science, Nizhny Novgorod 603950, Russia}
\affiliation{Nonlinear Physics Centre, Australian National University, Canberra ACT 2601, Australia}

\date{\today}

\begin{abstract}
We analyze the modulational instability of nonlinear Bloch waves in topological photonic lattices. In the initial phase of the instability development captured by the linear stability analysis, long wavelength instabilities and bifurcations of the nonlinear Bloch waves are sensitive to topological band inversions. At longer timescales, nonlinear wave mixing induces spreading of energy through the entire band and spontaneous creation of wave polarization singularities determined by the band Chern number. Our analytical and numerical results establish modulational instability as a tool to probe bulk topological invariants and create topologically non-trivial wave fields.
\end{abstract}

\maketitle

Topological photonic bands can be combined with appreciable mean-field nonlinear interactions in a variety of platforms~\cite{Topo_review,NonlinearTopo_review,review2}, including exciton-polariton condensates in structured microcavities~\cite{polariton_topo,Baboux2018}, waveguide arrays~\cite{Mukherjee_arxiv}, metasurfaces~\cite{Smirnova2019}, and ring resonators~\cite{topological_source}. These nonlinear topological photonic systems are of growing interest due to not only potential applications such as lasers and optical isolators, but also their ability to host novel effects with no analogue in electronic topological materials. For example, several previous studies have revealed the existence of self-localized wavepackets such as edge and bulk solitons~\cite{Ablowitz2014,Lumer2016,Kartashov2016,Leykam2016,Lumer2013,Poddubny2018,Marzuola_arxiv,Smirnova2019b}. \revision{However, as the size of a soliton is determined by its total power, they require fine-tuned excitation conditions to create, making experiments challenging. It is therefore timely to unveil novel phenomena that can emerge \emph{spontaneously} in nonlinear topological photonic systems, without requiring fine-tuning.}

In this paper we study the nonlinear dynamics of Bloch waves in topological bands, establishing their sensitivity to topological invariants such as the Chern number. We show that the generic phenomenon of modulational instability can lead to the spontaneous formation of wave fields characterized by non-trivial Chern numbers inherited from the linear bands. The mechanism is the energy-dependent parametric gain provided by the modulational instability~\cite{MI_review,NL_book,Kivshar1992,Engelhardt2015,Bardyn2016,Everitt2017,Nguyen2017,MI_gauge}, which enables population of a single band starting from a simple plane wave initial state. In addition to providing a simple way to sculpture novel structured light fields, the modulational instability enables measurement of bulk topological invariants of bosonic wave systems. This is usually challenging unless the band eigenstates are known a priori, time-consuming Bloch band tomography is performed~\cite{bardyn2014,aidelsburger2015,wimmer2017,tarnowski2019}, or the bulk-edge correspondence is employed~\cite{Poshakinskiy2015,chong,mittal2016}. 

We first characterize the short time dynamics of nonlinear Bloch waves using the linear stability analysis. The Bloch waves at high symmetry points of the Brillouin zone exhibit a familiar long wavelength instability in the presence of weak nonlinearities, but they become stable at a critical nonlinearity strength. This critical strength coincides with the bifurcation of a nonlinear Dirac cone~\cite{NDC}, where additional symmetry-breaking nonlinear Bloch waves emerge and their stability becomes sensitive to the band topology. Second, we use numerical simulations to study the modulational instability at longer propagation times. For weak nonlinearities the instability remains confined to the initially-excited band. Nonlinear wave mixing processes lead to the excitation of all the band's linear modes, imprinting the band's Chern number on the wave field's polarization~\cite{Foesel2017,leaky}. Interestingly, the polarization field converges to a quasi-equilibrium state well before the system thermalizes~\cite{Smerzi,Flach,Hafezi,Wu2019}. Thus, the topological properties of the band are imprinted on the modulational instability at small and large time and nonlinearity scales.

\begin{figure}

\includegraphics[width=\columnwidth]{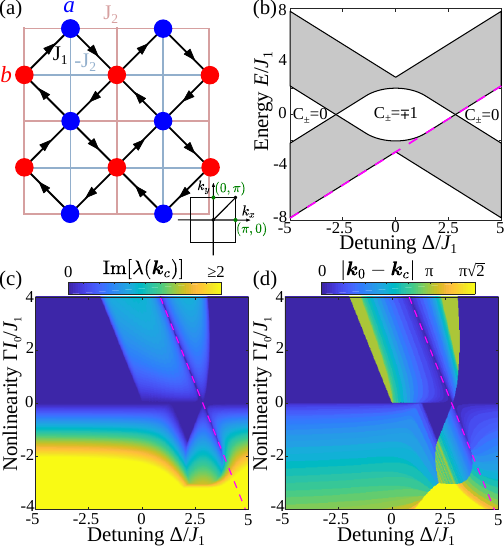}

\caption{Linear stability of the $\bs{k}_0=(\pi,0)$ nonlinear Bloch waves in the chiral-$\pi$-flux model. (a) Schematic of the lattice, consisting of two sublattices $(a,b)$ with detuning $\Delta$ and inter- (intra-) sublattice couplings $J_1$ ($J_2$). \revision{Inset: the Brillouin zone.} (b) Linear bands (shaded regions) \revision{characterized by Chern numbers $C_{\pm}$} as a function of $\Delta$, for $J_2=J_1/\sqrt{2}$. (c,d) Growth rate (c) and magnitude (d) of the most unstable perturbation wavevector $\bs{k}_c$. Purple dashed line in (c,d) marks the nonlinearity-induced gap closure at $\bs{k}=\bs{k}_0$,  \revision{and the nonlinear Bloch wave is stable in the dark blue areas.}} 

\label{fig:TFB_LSA}

\end{figure}

We consider a two-dimensional photonic lattice governed by the nonlinear Schr\"odinger equation,
\be 
i \partial_t \ket{\psi(\bs{r},t)} = (\hat{H}_L + \hat{H}_{N\!L}) \ket{\psi(\bs{r},t)}, \label{eq:SE}
\ee
where $t$ is the evolution variable (time or propagation distance), $\ket{\psi(\bs{r},t)}$ is the wave field profile, $\hat{H}_L$ and $\hat{H}_{N\!L}$ are linear and nonlinear parts of the Hamiltonian, and $\bs{r}=(x,y)$ indexes the lattice sites. We consider the chiral-$\pi$-flux model illustrated in Fig.~\ref{fig:TFB_LSA}(a). This is a two band tight binding model for a Chern insulator on a square lattice with two sublattices $a$ and $b$, i.e. $\ket{\psi} = (\psi_a,\psi_b)^T$, described by the Bloch Hamiltonian~\cite{FQH}
\begin{align}
&\hat{H}_L({\bm k}) = \bs{d}(\bs{k}) \cdot \bs{\hat{\sigma}}, \quad d_z = \Delta + 2 J_2 (\cos k_x - \cos k_y) \label{eq:TFB} \\
&d_x + i d_y = J_1 [ e^{-i\pi/4} (1+e^{i(k_y-k_x)}) + e^{i \pi/4} (e^{-ik_x} + e^{i k_y})],  \nonumber
\end{align}
where the wavevector $\bs{k}=(k_x,k_y)$ is restricted to the first Brillouin zone $k_{x,y} \in [-\pi,\pi]$, $\bs{\hat{\sigma}}=(\hat{\sigma}_x,\hat{\sigma}_y,\hat{\sigma}_z)$ are Pauli matrices acting on the sublattice (pseudospin) degree of freedom, $J_{1,2}$ are nearest and next-nearest neighbor hopping strengths, and $\Delta$ is a detuning between the sublattices, \revision{which we use to tune between trivial and non-trivial topological phases}. We will fix $J_2 = J_1/\sqrt{2}$, which enhances nonlinear effects by maximizing the band flatness~\cite{FQH}. For the nonlinear part of the Hamiltonian $\hat{H}_{N\!L}$ we consider an on-site nonlinearity of the form
\be 
\hat{H}_{N\!L} = \Gamma \mathrm{diag}[f(|\psi_a(\bs{r})|^2), f(|\psi_b(\bs{r})|^2)],
\ee
where $\Gamma$ is the nonlinear interaction strength and $f$ is the nonlinear response function, \revisiontwo{which describe the intensity-dependence of the site energies.}

The Bloch wave eigenstates of Eq.~\eqref{eq:TFB} form two energy bands $E_{\pm}(\bs{k})$, i.e. $\hat{H}_L(\bs{k})\ket{u_{\pm}(\bs{k})} = E_{\pm}(\bs{k}) \ket{ u_{\pm}(\bs{k})}$. Fig.~\ref{fig:TFB_LSA}(b) shows the spectrum of $\hat{H}_L$, \revision{$E_{\pm}^2 (\bm{k},\Delta)=4 J_{1}^2\left[ 1 + \cos k_{x} \cos k_{y} \right] + \left[ \Delta+ 2 J_{2}\left(\cos k_{x}-\cos k_{y}\right) \right]^2 $},
as a function of $\Delta/J_1$, which exhibits topological phase transitions at $\Delta/J_1 = 2 \sqrt{2}$ and $-2\sqrt{2}$, where the gap closes at \revision{one of the high symmetry points of the Brillouin zone $\bs{k}_0 = (\pi,0)$ and $(0,\pi)$, respectively. The phase transitions correspond to changes in the quantized Chern numbers $C_{\pm}$, which is computed as the integral of the Bloch waves' Berry curvature over the Brillouin zone~\cite{TI_review,Berry_review}.}

We focus on the nonlinear wave dynamics at the $\bs{k}_0 = (\pi, 0)$ high symmetry point. The linear Bloch wave can be continued as a nonlinear Bloch wave~\cite{Trager2006,NL_review} $\ket{\phi(\bs{r})} = (\sqrt{I}_0,0)^T e^{i \pi x}$ \revisiontwo{satisfying $(\hat{H}_L + \hat{H}_{N\!L})\ket{\phi} = E_{N\!L}\ket{\phi}$}, with energy $E_{N\!L} = \Delta - 4 J_2 + \Gamma f(I_0)$ bifurcating from the lower band when $\Delta < 4J_2$ and from the upper band when $\Delta > 4J_2$ [see purple line in Fig.~\ref{fig:TFB_LSA}(b)]. 

\revision{Performing the linear stability analysis, we consider the time evolution of a small perturbation to the steady state $\ket{\phi}$, $\ket{\psi(t)} = (\ket{\phi} + \ket{\delta\phi(t)}) e^{-i E_{N\!L}t}$. By linearizing the equation of motion Eq.~\eqref{eq:SE} about $\ket{\phi}$ we obtain an eigenvalue problem for the spectrum $\lambda(\bs{k})$ of perturbation modes $\ket{\delta \phi(t)} = \ket{u} e^{-i \lambda t} + \ket{v^*} e^{i \lambda^* t}$, \revisiontwo{which occur in complex conjugate pairs $\lambda, \lambda^*$ due to the particle-hole symmetry of the linearized eigenvalue problem}  (see Ref.~\cite{supp} for details).}

Perturbations with \revision{growth rate} $\mathrm{Im}(\lambda)>0$ are linearly unstable. For the nonlinear response function we consider pure Kerr nonlinearity $f(I) = I$, however the qualitative features of the perturbation spectrum are intensitive to the precise form of $f(I)$~\cite{supp}.

Figs.~\ref{fig:TFB_LSA}(c,d) plot the growth rate and wavevector $\bs{k}_c$ of the most unstable perturbation mode, 
\revisiontwo{i.e. the mode with the maximum growth rate $\mathrm{Im} [\lambda(\bs{k}_c)] = \max_{\bs{k}} \mathrm{Im} [\lambda(\bs{k})]$,} as a function of $\Delta$ and $\Gamma$. For weak nonlinearities $\Gamma$ we observe behaviour qualitatively similar to the scalar nonlinear Schr\"odinger equation: Bloch waves at the band edge exhibit a long wavelength instability under self-focusing nonlinearity, i.e. when $\Gamma m_{\mathrm{eff}} < 0$, where $m_{\mathrm{eff}} = \Delta - 4J_2$ is the wave effective mass at $\bs{k}_0$. Interestingly, a second long wavelength instability occurs for stronger nonlinearities in the vicinity of the stable line $\Gamma I_0/2 = -m_{\mathrm{eff}}$. This critical line occurs when the nonlinearity-induced potential closes the band gap and corresponds to a transition from an exponential instability at weak $\Gamma$ to an oscillatory instability at strong $\Gamma$.

To reveal the generic behavior in the vicinity of the critical line we consider the effective Dirac model obtained as a long wavelength expansion of Eq.~\eqref{eq:TFB}, i.e. $\bs{k} = \bs{k}_0 + \bs{p}$ with $|\bs{p}| \ll 1$~\cite{supp},
\be 
\hat{H}_D = J_1 \sqrt{2} ( p_x \hat{\sigma}_y - p_y \hat{\sigma}_x ) + (m_{\mathrm{eff}} + J_2[p_x^2 + p_y^2])\hat{\sigma}_z. \label{eq:dirac}
\ee
The quadratic $J_2[p_x^2+p_y^2]\hat{\sigma}_z$ term is required to reproduce the correct Chern numbers $C_{\pm} = \pm \frac{1}{2}(1-\mathrm{sgn}[J_2 m_{\mathrm{eff}}])$~\cite{Shen} and the main features of perturbation spectrum.

\begin{figure}

\includegraphics[width=\columnwidth]{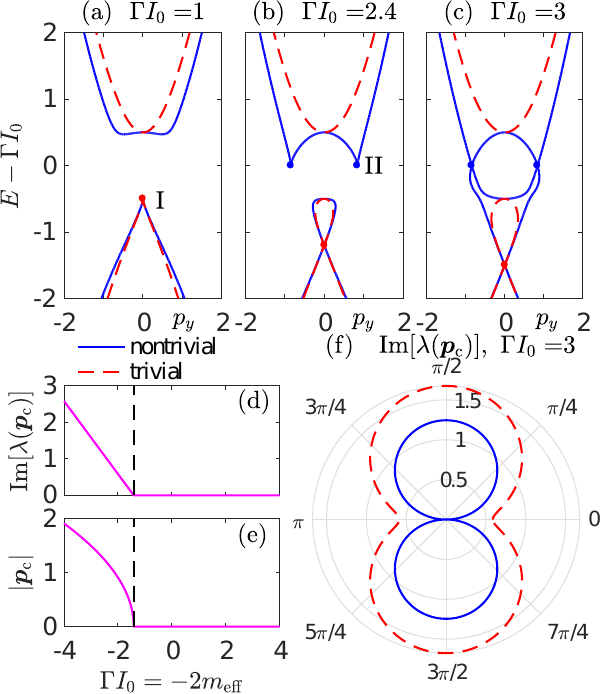}

\caption{(a-c) The transition in the nonlinear Bloch wave spectrum across the critical line in the non-trivial (solid blue; $m_{\mathrm{eff}}=-1/2$) and trivial (dashed red; $m_{\mathrm{eff}}=1/2$) phases of the effective Dirac model Eq.~\eqref{eq:dirac}. \revision{Colored dots mark special points in spectra with $p_y = 0, \pm p_{\text{II}}$.} (d,e) Growth rate (d) and magnitude (e) of the most unstable perturbation wavevector along the critical line. (f) Instability growth rate as a function of the polar angle measured from the symmetry-breaking nonlinear Bloch wave vector in the non-trivial (solid blue) and trivial (dashed red) phases.}

\label{fig:NLDE_LSA}

\end{figure}

The full nonlinear Bloch wave spectrum of Eq.~\eqref{eq:dirac} can be obtained analytically~\cite{supp} and is shown in Figs.~\ref{fig:NLDE_LSA}(a-c). We observe that the critical line coincides with the formation of a nonlinear Dirac cone at $\bs{k}_0$, \revision{with $|\bs{p}| = p_{\text{I}} = 0 $}~\cite{NDC}, i.e. a symmetry-breaking bifurcation \revision{[labeled I in Fig.~\ref{fig:NLDE_LSA}(a)]} of the nonlinear Bloch waves \revisiontwo{which produces the linear band crossing in Fig.~\ref{fig:NLDE_LSA}(b)}. At the bifurcation new modes $\ket{\phi(\bs{r})} = (\sqrt{I_a}e^{i \varphi},\sqrt{I_b})e^{i \pi x}$ emerge, \revision{where both sublattices have nonzero intensities satisfying $I_0 = I_a + I_b$ and the relative phase between them \revisiontwo{$\varphi$} forms a free parameter.} Moreover, in the non-trivial phase \revision{($C_{\pm}=\pm 1$, $|\Delta/J_1| < 2\sqrt{2}$)} an additional bifurcation \revision{[labeled II in Fig.~\ref{fig:NLDE_LSA}(b)]} occurs at higher intensities at $|\bs{p}| = \revision{p_{\text{II}}}= \sqrt{4 - \Delta/J_2}$, corresponding to $d_z(\bs{p}) = 0$. The new branches emerging from this bifurcation merge with the lower band as $\Gamma I_0$ is increased \revision{[see Fig.~\ref{fig:NLDE_LSA}(c)]}, producing a gapless nonlinear Bloch wave spectrum, while in the trivial phase \revision{($C_{\pm} = 0$, $|\Delta/J_1| > 2\sqrt{2}$)} the nonlinear Bloch wave spectrum remains gapped~\cite{supp}.

The nonlinear Dirac cone \revision{at $\bs{p} = 0$} occurs in both topological phases, but the modes' stability in the vicinity of the bifurcation point is sensitive to the linear band topology. For example, in both the tight binding and continuum models the critical stable line terminates abruptly in the trivial phase at $\Delta = \Delta_c =  4J_2 + \frac{J_1^2}{2 J_2}$, as shown in Fig.~\ref{fig:NLDE_LSA}(d,e). Beyond this critical detuning the most unstable wavevector is $|\bs{p}_c| = \sqrt{|\Gamma I_0 J_2 +J_1^2|/J_2^2}$; the length scale of the instability is dictated by the quadratic $J_2(p_x^2+p_y^2)$ term and vanishes in the usual linear Dirac approximation, which neglects $p_{x,y}^2$ terms. As a second example, Fig.~\ref{fig:NLDE_LSA}(f) shows the angular (directional) dependence 
of the maximal instability growth rate of the symmetry-breaking nonlinear Bloch wave. In the non-trivial phase the instability is strongly anisotropic~\cite{supp}, with wavevectors in the direction perpendicular to the direction of the pseudospin remaining stable, whereas in the trivial phase instabilities occur for all angles.

To understand these topological phase-dependent stability properties, we note that in the trivial phase the perturbation modes maintain a similar polarization to the nonlinear Bloch wave, enabling efficient nonlinear wave mixing and promoting instabilities. On the other hand, in the non-trivial phase the perturbation modes' polarization rotates away to the opposite pole of the Bloch sphere, reducing the strength of the nonlinear wave mixing due to poor spatial overlap between the nonlinear Bloch wave and the perturbation modes~\cite{footnote2}. While this difference may seem minor, it can play a critical role close to bifurcation points by lifting the degeneracy between the bands of perturbation modes. Thus, the modulational instability does not just depend on the energy eigenvalue dispersion, but is also sensitive to the geometrical properties of the Bloch waves, i.e. their polarization, and the band topology. This is our first key result.

Next, we carry out numerical simulations of Eq.~\eqref{eq:SE} to study the modulational instability beyond the initial linearized dynamics. To characterize the complex multi-mode dynamics, we compute the following observables: (i) The normalized real space participation number, which measures the fraction of strongly excited lattice sites,
\be 
P_{\revision{\textrm{r}}} = \frac{\mathcal{P}^2}{2N} \left( \sum_{\bs{r}}  |\psi_a(\bs{r})|^4 +  |\psi_b(\bs{r})|^4 \right)^{-1},
\ee
where $\mathcal{P} = \sum_{\bs{r}} \braket{\psi(\bs{r})|\psi (\bs{r})}$ is the total power. (ii) The Fourier space participation number $P_{\revision{\textrm{k}}}$, which measures similarly the fraction of excited Fourier modes. (iii) The field polarization direction $\hat{\bs{n}}_{\psi}(\bs{k}) = \bra{\psi(\bs{k})} \bs{\hat{\sigma}} \ket{\psi(\bs{k})}/\braket{\psi(\bs{k})|\psi(\bs{k})}$, which exhibits singularities sensitive to the band topology. We average these observables over an ensemble of small random initial perturbations to the nonlinear Bloch wave. The average polarization $\langle \hat{\bs{n}}_{\psi}(\bs{k}) \rangle$ in general describes a mixed state with $n^2_{\psi} = \langle \hat{\bs{n}}_{\psi}(\bs{k}) \rangle \cdot \langle \hat{\bs{n}}_{\psi}(\bs{k}) \rangle < 1$. When $n_{\psi}^2 > 0$ for all $\bs{k}$, i.e. the ``purity gap'' $\mathrm{min}_{\bs{k}} (n_{\psi}^2)$ remains open, the wave field is characterized by a quantized Chern number~\cite{Hu2016,Bardyn2013,Budich2015}. 

\begin{figure}

\includegraphics[width=0.8\columnwidth]{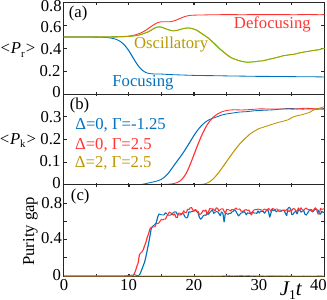}

\caption{Long time instability dynamics in the different instability regimes: focusing exponential (blue), defocusing exponential (red), and oscillatory instability (brown). (a) Real space participation number. (b) Fourier space participation number. (c) Purity gap.}

\label{fig:BHZ_long}

\end{figure}

Fig.~\ref{fig:BHZ_long} illustrates the dynamics of the $\bs{k}_0 = (\pi,0)$ nonlinear Bloch wave with initial intensity $I_0 = 1$, when each lattice site is subjected to a random perturbation with amplitude not exceeding $0.01\sqrt{I_0}$. We use saturable nonlinearity of the form $f(I) = 2I/(1+I)$, which takes into account the inevitable saturation of nonlinear response at high intensities, a system size of $N = 32 \times 32$ unit cells with periodic boundary conditions~\cite{footnote1}, and average observables over 100 initial perturbations. We consider parameters corresponding to different instability regimes: exponential focusing, exponential defocusing, and oscillatory defocusing. 

The focusing instability generates a collection of localized solitons, resulting in a decrease in $\langle P_{\revision{\textrm{r}}} \rangle$ in Fig.~\ref{fig:BHZ_long}(a). On the other hand, the defocusing nonlinearity spreads energy over both sublattices, resulting in a small increase in $\langle P_{\revision{\textrm{r}}} \rangle$. In all cases $\langle P_{\revision{\textrm{k}}} \rangle$ increases due to other Fourier modes being populated via nonlinear wave mixing. For the exponential instabilities this is accompanied by the purity gap opening and emergence of a well-defined Chern number corresponding to the band Chern number. Interestingly, the purity gap opens before $\langle P_{\revision{\textrm{r},\textrm{k}}}\rangle$ reach a steady state. Under the oscillatory instability the purity gap remains negligible due to competition between pairs of instability modes with the same growth rates. \revision{Further details of the propagation dynamics in these different regimes may be found in Ref.~\cite{supp}.}

\begin{figure}

\includegraphics[width=\columnwidth]{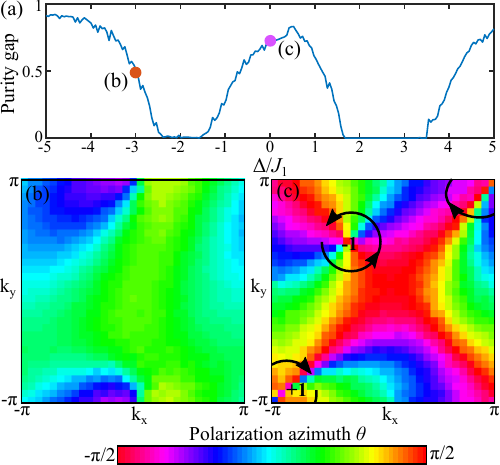}
\caption{(a) Purity gap at time $t=t_f = 40/J_1$ as a function of $\Delta$. (b,c) Field polarization textures at $t_f$ in the (b) trivial ($\Delta = -3J_1$) and (c) non-trivial ($\Delta = 0$) phases. The Chern number is obtained by summing the charges of the polarization azimuth vortices (indicated by arrows) weighted by $\mathrm{sgn}(n_y)$ at the vortex core (indicated by $\pm 1$)~\cite{Foesel2017}.}

\label{fig:Delta}

\end{figure}

To explore the emergence of a purity gap in more detail, we present in Fig.~\ref{fig:Delta}(a) its value at $t=t_f=40J_1$ as a function of $\Delta$, which tunes between the trivial and non-trivial phases~\cite{FQH}. For $\Delta>0$ we observe good correspondence with the results of the linear stability analysis in Fig.~\ref{fig:TFB_LSA}: the size of the purity gap follows the instability growth rate, and the purity gap vanishes when the Bloch wave is linearly stable or exhibits an oscillatory instability, because the polarization becomes sensitive to the initial random perturbation. The purity gap also closes at $\Delta \approx -2 J_1$ despite no change in the fastest growing instability mode, corresponding to a nonlinearity-induced closure of the band gap at $\bs{k} = (0,\pi)$.

While the trivial and non-trivial phases exhibit similar purity gaps, their differing topology can be observed by measuring the field polarization $\langle \hat{\bs{n}}_{\psi}(\bs{k})\rangle = (n_x,n_y,n_z)$ at long times, as illustrated in Fig.~\ref{fig:Delta}(b,c). \revision{In experiments $\langle \hat{\bs{n}}_{\psi}(\bs{k})\rangle$ can be readily obtained by measuring the field's sublattice (spin) components in Fourier space; $n_z(\bs{k}) = (|\psi_a|^2 - |\psi_b|^2)/(|\psi_a|^2+|\psi_b|^2)$ is the relative population imbalance between the two sublattices at $\bs{k}$, while $n_{x,y}$ depend on the relative phase between the two sublattices.} Employing the method of Ref.~\cite{Foesel2017}, the Chern number can hence be obtained by summing the charges of the phase singularities of the polarization azimuth $\theta = \frac{1}{2} \tan^{-1}(n_x/n_z)$ weighted by $\mathrm{sgn}(n_y)$. In the trivial phase (large $\Delta$) the field is predominantly localized to a single sublattice, such that $n_z$ remains nonzero and there are no phase singularities in $\theta$; hence $C=0$. In the non-trivial phase $\langle \hat{\bs{n}}_{\psi}(\bs{k})\rangle$ spans the entire Bloch sphere, corresponding to a pair of opposite charge phase singularities with opposite weights $\mathrm{sgn}(n_y) = \pm 1$, and hence $C=1$. Thus, the long time instability dynamics can be used to measure the band Chern number. This is our second important finding.

In conclusion, we have studied how the modulational instability of nonlinear Bloch waves can be used to probe band topology. The linear stability spectrum describing the short time instability dynamics exhibits bifurcations and a re-emergence of stability which are sensitive to topological band inversions. At longer evolution times nonlinear wave mixing can populate an entire band, enabling the spontaneous creation of topologically non-trivial wave fields from simple plane wave initial states. Since the timescales involved are shorter than the wave thermalization time, these effects should be experimentally observable in nonlinear waveguide arrays~\cite{Mukherjee_arxiv}, Bose-Einstein condensates in optical lattices~\cite{Everitt2017,Nguyen2017}, or exciton-polariton condensates~\cite{polariton_topo,Baboux2018}. While we focused on the chiral-$\pi$-flux model, we have observed similar behaviour in other topological tight binding models. Lattices with a larger band flatness typically exhibit emergence of a purity gap and well-defined Chern number for a wider range of nonlinearity strengths. It will be interesting to generalize our findings to periodically-driven Floquet systems such as the nonlinear waveguide array employed in Ref.~\cite{Mukherjee_arxiv}, where perfectly flat topological bands have been demonstrated.

\acknowledgements
We thank Mikael Rechtsman for illuminating discussions. This research was supported by the Institute for Basic Science in Korea (IBS-R024-Y1, IBS-R024-D1), the Australian Research Council Early Career Researcher Award (DE190100430), and the Ministry of Education, Science, and Technological Development of the Republic of Serbia. Theoretical analysis of the continuum model was supported by the Russian Science Foundation (Grant No. 20-72-00148).


\widetext

\setcounter{equation}{0}
\setcounter{figure}{0}
\setcounter{table}{0}
\makeatletter
\renewcommand{\theequation}{S\arabic{equation}}
\renewcommand{\thefigure}{S\arabic{figure}}
\renewcommand{\bibnumfmt}[1]{[S#1]}
\renewcommand{\citenumfont}[1]{S#1}

\begin{center}
\textbf{\Large Supplementary Material for \\
``Probing band topology using modulational instability''}

{Daniel Leykam}$^{1,2,*}$, {Ekaterina Smolina$^{3}$, {Aleksandra Maluckov}$^{1,4}$, {Sergej Flach}$^{1,2}$, and Daria Smirnova$^{3,5}$}

{\it $^1${Center for Theoretical Physics of Complex Systems, Institute for Basic Science, Daejeon 34126, Korea}

$^2${Basic Science Program, Korea University of Science and Technology, Daejeon 34113, Korea}

$^3${Institute of Applied Physics, Russian Academy of Science, Nizhny Novgorod 603950, Russia}

$^4${P$^{*}$ Group, Vin\v{c}a Institute of Nuclear Sciences, University of Belgrade, P.O. Box 522, 11001 Belgrade, Serbia}

$^5$Nonlinear Physics Center, Research School of Physics, Australian National University, \\Canberra ACT 2601 Australia
}
\end{center}

\section{\large{Linear Bloch wave spectrum}}

The linear Hamiltonian of the two-band Chern insulator model considered in the main text is
\begin{equation}
    \hat{H}_L({\bm k}) = \bs{d}(\bs{k}) \cdot \bs{\hat{\sigma}}, \quad d_z = \Delta + 2 J_2 (\cos k_x - \cos k_y) \quad d_x + i d_y = J_1 [ e^{-i\pi/4} (1+e^{i(k_y-k_x)}) + e^{i \pi/4} (e^{-ik_x} + e^{i k_y})].  \nonumber
\end{equation}
Its spectrum forms two bulk bands $E_{\pm} = \pm |\bs{d}| = \pm \sqrt{d_x^2 + d_y^2 + d_z^2}$. Explicitly, 
\begin{equation}
E_{\pm}({\bm k}, \Delta)=\pm \sqrt{4 J_{1}^2\left[ 1 + \cos k_{x} \cos k_{y}  \right] + \left( \Delta+2 J_{2}\left(\cos k_{x}-\cos k_{y}\right) \right)^2}\:. \label{eq:spectrum}
\end{equation}
The Bloch wave eigenstates $\hat{H}_L(\bs{k}) \ket{u_{\pm}(\bs{k})} = E_{\pm}(\bs{k}) \ket{u_{\pm}(\bs{k})}$ are characterized by a topological invariant, the Chern number~\cite{sTI_review}
\begin{equation}
C_{\revision{\pm}} = \frac{1}{2\pi} \iint_{BZ} \mathcal{F}_{\revision{\pm}} (\bs{k}) \revision{ \text{d}^2 {\bs k} } , \label{eq:chern_number}
\end{equation}
where the integral is computed over the Brillouin zone in the reciprocal space, and 
\begin{equation}
\mathcal{F}_{\revision{\pm}} (\bs{k}) = i \left[ \braket{\partial_{k_x}u_{\pm} | \partial_{k_y}u_{\pm}} - \braket{\partial_{k_y}u_{\pm} | \partial_{k_x} u_{\pm}}\right]
\end{equation}
is the Berry curvature. In two band systems the Bloch functions $\ket{u_{\pm}(\bs{k})}$ can be expressed as a real unit polarization vector $\hat{\bs{n}}$ on the surface of the Bloch sphere. To see this, we use the Pauli matrix parameterization of the Bloch Hamiltonian $\hat{H}_L = \bs{d}(\bs{k}) \cdot \bs{\hat{\sigma}}$ and multiply the eigenvector equation below Eq.~\eqref{eq:spectrum} by $\bra{u_{\pm}(\bs{k})}$, yielding
$$
\bs{d} \cdot \braket{u_{\pm}|\bs{\hat{\sigma}}|u_{\pm}} = E_{\pm} = \pm |\bs{d}| \Rightarrow \bs{\hat{n}}_{\pm} \equiv \braket{u_{\pm}|\bs{\hat{\sigma}}|u_{\pm}} = \pm \bs{d} / |\bs{d}|.
$$
Moreover, using the $\bs{d}$ vector parameterization, the Berry curvature takes the simplified form $\mathcal{F}_{\pm} = \pm \frac{1}{2}\frac{\bs{d}}{|\bs{d}|^3}$~\cite{sBerry_review}. It follows that the Berry curvature can be written in terms of the polarization vector as
\begin{equation}
\mathcal{F}_{\revision{\pm}} (\bs{k}) = -\frac{1}{2} \hat{\bs{n}}_{\revision{\pm}} \cdot [ (\partial_{k_x} \hat{\bs{n}}_{\revision{\pm}} ) \times (\partial_{k_y} \hat{\bs{n}}_{\revision{\pm}} )],
\end{equation} 
with the Chern number counting the number of times $\hat{\bs{n}}_{\pm}$ covers the unit sphere~\cite{sTI_review}. We note that the interpretation of $C_{\revision{\pm}} $ in terms of the wave polarization field can also be generalized to multi-band systems~\cite{sFoesel2017}.

\begin{figure}[h]
\centering
{\includegraphics[width=0.5\textwidth]{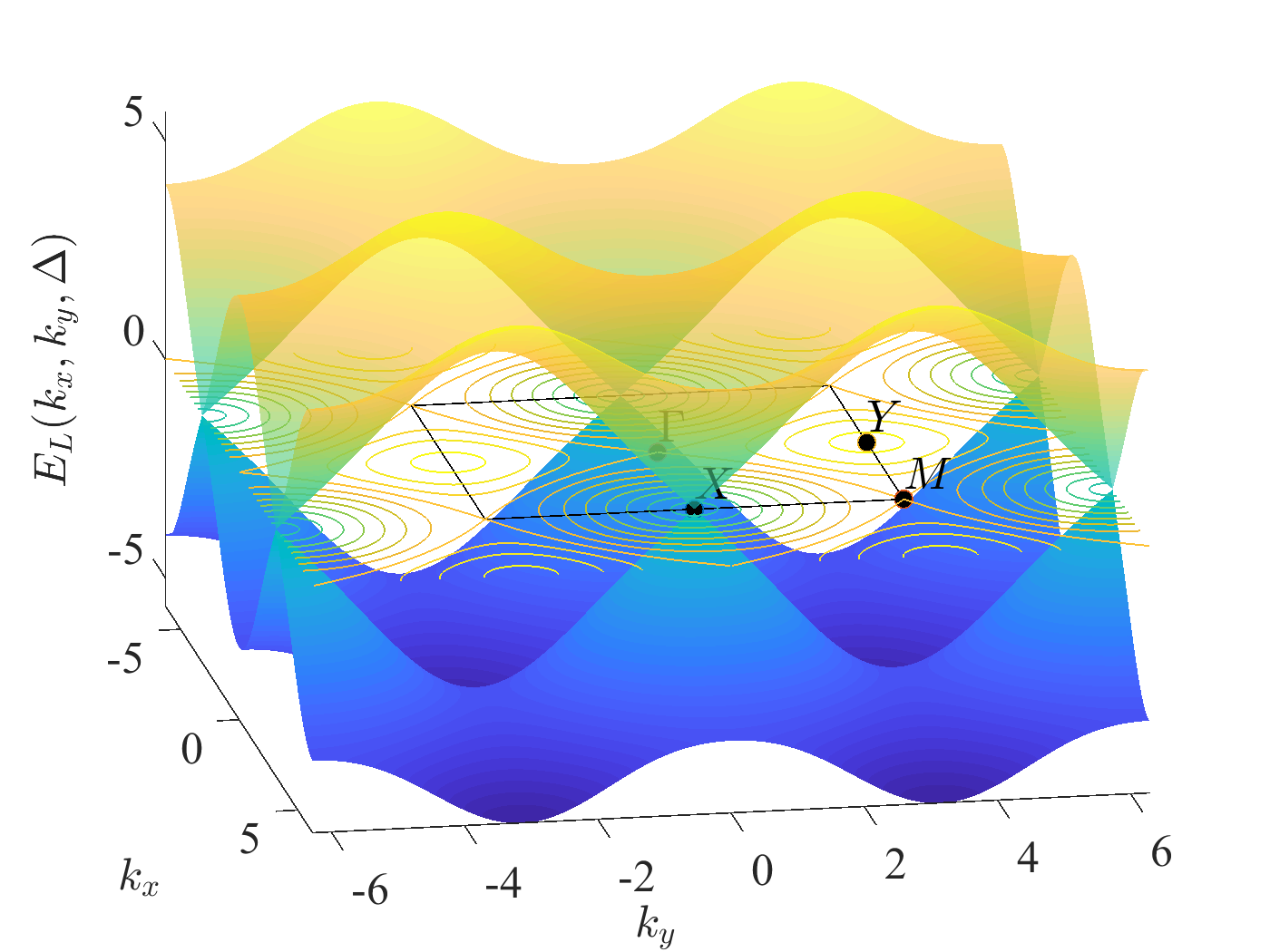}}
\caption{\revision{3D representation of the dispersion $E(k_x,k_y)$ for a square $\pi$-flux lattice with parameters $J_1=1,~J_2=J_1/\sqrt{2},~\Delta = 2 \sqrt{2} J_1$. Black dots mark the high-symmetry points of the first Brillouin zone traced with a black square.}}
\label{fig:figS_dispn}
\end{figure}

\revision{The quantized Chern number can only change at topological transitions where the band gap closes. In our two-band Chern insulator model the topological transition occurs when the gap between two bands closes and reopens at $\Delta = \pm 4 J_2$. This critical detuning separates phases with zero ($|\Delta|>4|J_2|$) and nonzero ($|\Delta|<4|J_2|$) Chern numbers, as depicted in Fig.~1(b) in the main text. The bandgap closure takes place at the high-symmetry points of the Brillouin zone:  $\boldsymbol{k}_0=(\pi,0)$ [for $\Delta J_2 > 0$, $\Delta = 4 J_2$] or 
$(0,\pi)$ [for $\Delta J_2 < 0$, $\Delta = - 4 J_2$], where the two
bands touch forming a spectral degeneracy, shown in Fig.~\ref{fig:figS_dispn}.
}

\section{\large Linear stability analysis}

Here we consider the linear stability of nonlinear Bloch waves in a generic tight binding lattice described by the nonlinear evolution equation
\begin{equation} 
i \partial_t \ket{\psi(\bs{r},t)} = (\hat{H}_L + \hat{H}_{NL}) \ket{\psi(\bs{r},t)}.
\end{equation} 
We assume that the nonlinear part of the Hamiltonian $\hat{H}_{NL}$ is a diagonal matrix with real elements dependent only on the local on-site intensity, i.e. $\hat{H}_{NL} = \Gamma \mathrm{diag}[f(|\psi_a|^2), f(|\psi_b|^2),...]$, where $f$ describes the intensity-dependent nonlinear frequency shift and $a,b,...$ indexes the sublattice degree of freedom. The linear part of the Hamiltonian $\hat{H}_L$ can be expanded in real space as
\begin{equation} 
\hat{H}_L \ket{\psi(\bs{r})} = \sum_{\bs{\delta}} \hat{C}(\bs{\delta})\ket{\psi(\bs{r}+\bs{\delta})},
\end{equation} 
where summation $\bs{\delta}$ is over neighbouring unit cells. Transforming to Fourier space, $\ket{\psi(\bs{r})} = \sum_{\bs{k}} \ket{\psi(\bs{k})} e^{i \bs{k}\cdot\bs{r}}$, we obtain the Bloch Hamiltonian
\begin{equation} 
\hat{H}(\bs{k}) \ket{\psi(\bs{k})} = \left( \sum_{\bs{\delta}} \hat{C}(\bs{\delta}) e^{i \bs{k}\cdot\bs{\delta}} \right) \ket{\psi(\bs{k})}.
\end{equation} 
Note that under the Fourier transform $\hat{H}_L^* \ket{\psi(\bs{r})} \rightarrow \hat{H}^*(-\bs{k}) \ket{\psi(\bs{k})}$.

To perform the linear stability analysis we consider small perturbations about some nonlinear steady state $\ket{\phi(\bs{r})}$ with energy $E$, i.e. $\ket{\psi(\bs{r},t)} = (\ket{\phi(\bs{r})} + \ket{\delta \phi(\bs{r},t)})e^{-i E t}$. First, by Taylor expansion of the diagonal nonlinear term and neglecting terms quadratic in the perturbation, we obtain a linearised evolution equation for the perturbation,
\begin{equation} 
(i \partial_t + E)\ket{p(\bs{r})} = \hat{H}_L \ket{p(\bs{r})} + \Gamma \sum_{j=a,b,...} \left[(f(|\phi_j|^2) + f^{\prime}(|\phi_j|^2)|\phi_j|^2) p_j(\bs{r}) +f^{\prime}(|\phi_j|^2) \phi_j^2 p_j^* (\bs{r}) \right] \ket{j},
\end{equation} 
The solution to this set of coupled first order linear differential equations can be expanded in terms of exponential functions as $\ket{\delta \phi(\bs{r},t)} = \ket{w(\bs{r})} e^{-i \lambda t} + \ket{v^*(\bs{r})} e^{i \lambda^* t}$. We collect terms with the same time dependence to obtain the eigenvalue problem 
\begin{align}
\lambda \ket{w(\bs{r})} &= (\hat{H}_L - E)\ket{w(\bs{r})} + \Gamma \sum_{j=a,b,...} \left[ (f(|\phi_j|^2) + f^{\prime}(|\phi_j|^2)|\phi_j|^2) w_j(\bs{r}) +  f^{\prime}(|\phi_j|^2)\phi_j^2 v_j(\bs{r}) \right]\ket{j}, \\
\lambda \ket{v(\bs{r})} &= -(\hat{H}_L^* - E)\ket{v(\bs{r})} - \Gamma \sum_{j=a,b,...} \left[ (f(|\phi_j|^2) + f^{\prime}(|\phi_j|^2)|\phi_j|^2) v_j(\bs{r}) + f^{\prime}(|\phi_j|^2)\phi_j^{*2} w_j(\bs{r}) \right] \ket{j}.
\end{align} 
Now we assume that steady state is a nonlinear Bloch wave such that $\ket{\phi(\bs{r})} = \ket{\phi} e^{i \bs{k_0}\cdot\bs{r}}$. Fourier transforming the above equations, there is coupling between perturbation fields $\ket{w(\bs{k})}$ and $\ket{v(\bs{k} - 2\bs{k}_0)}$. We obtain the coupled equations
\begin{align} 
\lambda \ket{w(\bs{k} + \bs{k}_0)} &= (\hat{H}(\bs{k}_0 + \bs{k}) - E) \ket{w} + \Gamma \sum_{j=a,b,...} \left[ (f(|\phi_j|^2) + f^{\prime}(|\phi_j|^2) |\phi_j|^2] ) w_j + f^{\prime}(|\phi_j|^2) \phi_{j}^2 v_j \right] \ket{j}, \\
\lambda \ket{v(\bs{k} - \bs{k}_0)} &= -(\hat{H}^*(\bs{k}_0 - \bs{k}) - E) \ket{v} - \Gamma \sum_{j=a,b,..} \left[ (f(|\phi_j|^2) + f^{\prime}(|\phi_j|^2)|\phi_j|^2] ) w_j + f^{\prime}(|\phi_j|^2) \phi_{j}^{*2} w_j \right] \ket{j}.
\end{align}
This eigenvalue problem has a built-in particle hole symmetry: eigenvalues $\lambda$ must occur in complex conjugate pairs. Real $\lambda$ correspond to stable perturbation modes, purely imaginary $\lambda$ result in exponential instabilities, and complex $\lambda$ correspond to oscillatory instabilities.

It is instructive to first consider the limit of weak nonlinearity $\Gamma \ll 1$. In this case, we can treat the terms arising due to the nonlinearity as a weak perturbation, $\lambda = \lambda_0 + \Gamma \lambda_1$. The unperturbed eigenvalue problem is
\begin{equation}
    \lambda_0 \left( \begin{array}{cc} \ket{w} \\ \ket{v} \end{array} \right) = \left(\begin{array}{cc} \hat{H}(\bs{k}_0 + \bs{k}) - E & 0 \\ 0 & -\hat{H}^*(\bs{k}_0 - \bs{k}) + E \end{array}\right) \left( \begin{array}{cc} \ket{w} \\ \ket{v} \end{array} \right).
\end{equation}
The eigenvalues are $\lambda_0 = E_n(\bs{k}_0+\bs{k}) - E, -E_n(\bs{k}_0+\bs{k}) + E$, with Bloch wave eigenvectors $(\ket{u_n(\bs{k}_0+\bs{k})},0)^T$, $(0, \ket{u_n^*(\bs{k}_0-\bs{k})})^T$, respectively. Assuming non-degenerate eigenvalues, the first order corrections due to the terms arising from the nonlinearity are
\begin{equation} 
\lambda_1 = \sum_{j=a,b,...} [f(|\phi_j|^2) + f^{\prime}(|\phi_j|^2)|\phi_j|^2] |\braket{u_n(\bs{k}_0 \pm \bs{k})|j}|^2 = g_{\mathrm{eff}}. \label{eq:geff}
\end{equation}
Thus, in this perturbative limit the perturbation modes maintain the same polarization as the lattice's Bloch waves, and their energy shifts are proportional to the squared overlap between linear Bloch waves and the nonlinear stationary state, which justifies the arguments used in the main text.

In two band tight binding models the Bloch Hamiltonian can be parameterized using the Pauli matrices as $\hat{H}(k) = \bs{d}(k) \cdot \hat{\bs{\sigma}}$, where $\bs{d}(k)$ is a real 3 component vector. We obtain the explicit matrix form of the linear stability equations,
\begin{equation} 
\lambda \left( \begin{array}{c} \ket{w} \\ \ket{v} \end{array} \right) = \left( \begin{array}{cc} \bs{d}(\bs{k}_0+\bs{k})\cdot\hat{\bs{\sigma}} - E + \Gamma \sum_j (f_j + f_j^{\prime}|\phi_j|^2) \ket{j}\bra{j} & \Gamma \sum_j f_j^{\prime} \phi_{j}^{2} \ket{j}\bra{j} \\ -\Gamma \sum_j f_j^{\prime} \phi_{j}^{*2} \ket{j}\bra{j} & - \bs{d}(\bs{k}_0-\bs{k})\cdot \hat{\bs{\sigma}}^* + E - \Gamma \sum_j (f_j + f_j^{\prime}|\phi_j|^2) \ket{j}\bra{j} \end{array} \right) \left( \begin{array}{c} \ket{w} \\ \ket{v} \end{array} \right),
\end{equation}
where $f_j = f(|\phi_j|^2)$ and $f_j^{\prime} = f^{\prime}(|\phi_j|^2)$. 

For the case of a nonlinear Bloch wave with intensity $I_0$ localized to the $a$ sublattice analyzed in the main text, we have $\ket{\phi} = (\sqrt{I_0},0)$, $d_{x,y}(\bs{k}_0) = 0$, $d_z(\bs{k}_0) = \Delta - 4J_2$, and $E = d_z(\bs{k}_0) + \Gamma f(I_0)$, and the $\bs{k}=0$ eigenvalue problem takes the particularly simple form
\begin{equation} 
\lambda \left( \begin{array}{c} w_a \\ w_b \\ v_a \\ v_b \end{array}\right) = \left( \begin{array}{cccc} \Gamma f^{\prime} I_0 & 0 & \Gamma f^{\prime} I_0 & 0 \\ 0 &  -2d_z(\bs{k}_0) - \Gamma f & 0 & 0 \\ - \Gamma f^{\prime} I_0 & 0 & -\Gamma f^{\prime} I_0 & 0 \\ 0 & 0 & 0 & 2 d_z(\bs{k}_0) + \Gamma f \end{array} \right) \left( \begin{array}{c} w_a \\ w_b \\ v_a \\ v_b \end{array}\right),
\end{equation}
yielding $\lambda = 0,0, \pm [2 d_z (\bs{k}_0) + \Gamma f(I_0)]$ and a fourfold degeneracy when $\Gamma f(I_0)/2 = -d_z(\bs{k}_0)$, i.e. when the nonlinear energy shift on the $a$ sublattice is sufficient to close the band gap. Complex instability eigenvalues emerge beyond this threshold intensity.

\begin{figure}[h]
\centering
{\includegraphics[width=0.6\textwidth]{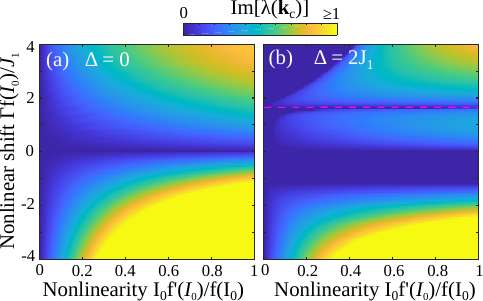}}
\caption{Growth rate of the most unstable perturbation wavevector $\bs{k}_c$ as a function of the nonlinear frequency shift $\Gamma f(I_0)$ and the relative nonlinear wave mixing strength $I_0 f^{\prime}(I_0)/f(I_0)$, for the $\Delta = 0$ (a) and $\Delta = 2J_1$ (b) lattices. Purple dashed line in (b) marks the nonlinearity-induced gap closure at $\bs{k} = \bs{k}_0$, which is only sensitive to the nonlinear frequency shift.}
\label{fig:LSA_response}
\end{figure}

The threshold intensity is independent of $f^{\prime}(I_0)$, i.e. the precise form of the nonlinear response function. For example, Fig.~\ref{fig:LSA_response} shows the dependence of the maximal growth rate on $\Gamma f(I_0)$ and the ratio $r = I_0 f^{\prime}(I_0)/f(I_0)$. The former describes the nonlinearity-induced change to the sublattice depths, while the latter describes the relative strength of nonlinearity-induced mixing between different perturbation wavevectors. $r$ interpolates between a completely saturated nonlinearity ($r = 0$, corresponding to effectively linear propagation dynamics), and pure Kerr nonlinearity ($r=1$). We observe that tuning $r$ does not affect the position of the critical stable line, and merely rescales instability growth rate. Thus, the linear stability eigenvalue spectra for the pure Kerr and saturable cases are qualitatively similar, exhibiting the same re-emergence of stability.

\section{\large Nonlinear Dirac model}

In this section we examine the dispersion and linear stability of the nonlinear Bloch waves in the vicinity of the high-symmetry point with in-plane wave vector ${\bm{k}}_0=(\pi,0)$, which can be described by an effective continuum model. In the vicinity of ${\bm{k}}_0$, ${\bm{k}}={\bm{k}}_0+{\bm{p}}$, the series expansion in the Bloch Hamiltonian $\hat{H}_{L} ({\bm{k}}) $ for $|{\bm{p}}| \ll 1$ leads to the Dirac-like Hamiltonian
\begin{subequations}
\begin{align}
\hat{H}_{D}({\bm{k}}_0 = (0,\pi)) &=-J_{1} \sqrt{2}\left( -p_{x} \hat{\sigma}_{x} + p_{y} \hat{\sigma}_{y}\right)+\left(\Delta-4 J_{2}+J_{2}\left[p_{x}^{2}+p_{y}^{2}\right]\right) \hat{\sigma}_{z} \:,  \\ 
\hat{H}_{D}({\bm{k}}_0 = (\pi,0)) & =-J_{1} \sqrt{2}\left( - p_{x} \hat{\sigma}_{y}+p_{y} \hat{\sigma}_{x}\right)+\left(\Delta-4 J_{2}+J_{2}\left[p_{x}^{2}+p_{y}^{2}\right]\right) \hat{\sigma}_{z}. \label{eq:H_D_pi_2}
\end{align}
\end{subequations}
The corresponding evolution equations including the local Kerr nonlinearity $f(I) = I$ can be formulated in the real space in terms of spatial derivatives by substituting $p_{x,y} = - i \partial_{x,y} $ into Eq.~\eqref{eq:H_D_pi_2}:
\begin{equation} 
\label{eq:main_sys_der}
i\partial_t \bm{\psi} \! = \left( \!{\begin{array}{*{22}{c}}
{\Delta-4 J_{2}-J_{2}\left[\partial_{x}^{2}+\partial_{y}^{2}\right] + \Gamma |\psi_1^2|} & {J_{1}\sqrt{2}\left( i\partial_y - \partial_x \right) }\\
{J_{1}\sqrt{2}\left( i \partial_y + \partial_x \right)}  & {-\Delta+4 J_{2}+J_{2}\left[\partial_{x}^{2}+\partial_{y}^{2}\right]+ \Gamma |\psi_2^2|}
\end{array}} \! \right) \bm{\psi}.
\end{equation}

\subsection{Nonlinear dispersion of bulk modes} 

We search for the solution of \eqref{eq:main_sys_der} in the form of weakly nonlinear Bloch waves:
\begin{equation} 
\label{eq:Bws}
\left( \!{\begin{array}{*{20}{c}}
{ \psi_1 } \\
{ \psi_2}
 \end{array}} \!  \right) =
 \left( \!{\begin{array}{*{20}{c}}
{ A} \\
{ B}
\end{array}} \!  \right) e^{-i E t + i p_x x + i p_y y}. 
\end{equation}
Plugging the spinor~\eqref{eq:Bws} into Eq.\eqref{eq:main_sys_der} results in the system of equations for the amplitudes $A$ and $B$
\begin{equation}
\label{eq:e_vect}
\begin{cases}
    {(-E+\Delta-4 J_{2}+J_{2}\left[p_{x}^{2}+p_{y}^{2}\right] + \Gamma |A^2|)A-J_{1}\sqrt{2}\left( p_y + i p_x \right)B = 0} \:, \\
       {(E+\Delta-4 J_{2}+J_{2}\left[p_{x}^{2}+p_{y}^{2}\right] - \Gamma |B^2|)B+J_{1}\sqrt{2}\left( p_y - i p_x \right)A  = 0 }\:,
 \end{cases}
\end{equation}
where the wave vector $ {\bm p} = (p_x,p_y) = p(\cos \theta, \sin \theta)$ can be defined in the polar coordinate system, $p_y + i p_x = i p e^{-i\theta}$.

Denoting the total wave intensity $|A|^2+|B|^2= I_0$, we first find the solutions for the lower and upper bands at the zero wave vector $\bm{p}=0$:
\begin{gather} 
\label{eq:Eu0}
  A^{(0)} =0, \quad |B^{(0)}|^2 =I_0, \quad E_2^{(0)} =- \Delta+4 J_{2}  + \Gamma|B^{(0)}|^2 = - \Delta+4 J_{2}  + \Gamma I_0 , \\
 B^{(0)} =0,  \quad |A^{(0)}|^2 =I_0, \quad E_1^{(0)} = \Delta-4 J_{2}  + \Gamma |A^{(0)}|^2 = \Delta-4 J_{2}  + \Gamma I_0 .
\end{gather}
At the intensities above the critical value $ I_0 \ge \pm  2 \frac{\left( \Delta - 4 J_2 \right)}{\Gamma}$, we get the additional doubly degenerate solution  
\begin{equation} 
|A^{(0)}|^2 = \dfrac{I_0}{2}  -  \dfrac{ \Delta-4 J_{2}  }{\Gamma}, \quad |B^{(0)}|^2 = \dfrac{I_0}{2}  + \dfrac{ \Delta-4 J_{2} }{\Gamma}, \quad E_3^{(0)} =   \frac { \Gamma I_0}{2}
\end{equation}
with the eigenvectors: 
\begin{equation} \label{eq:spinor}
\left( \!{\begin{array}{*{20}{c}}
{ A^{(0)}  } \\
{ B^{(0)} }
 \end{array}} \!  \right) =
 \dfrac{1}{\sqrt{2}}\left( \!{\begin{array}{*{20}{c}}
{e^{i \varphi} \sqrt{I_0-\dfrac{2\left( \Delta - 4 J_2 \right)}{\Gamma}}} \\
{ \pm \sqrt{I_0 +\dfrac{2\left( \Delta - 4 J_2 \right)}{\Gamma}}}
\end{array}} \!  \right)\:,
\end{equation}
where $\varphi$ is an arbitrary phase depending on which direction we approach the degeneracy point $\bm{p}=0$. Specifically, this phase uncertainty is lifted 
if we consider the limit transition to the point $\bm{p}=0$ along different directions ${\bm p}_0 (p_0 \rightarrow 0 ) = (p_{0x}, p_{0y} ) = p_0(\cos \theta_0, \sin \theta_0)$. According to \eqref{eq:e_vect}, the phase shift in spinor~\eqref{eq:spinor} is given by $\varphi = \pi/2 - \arctan(p_{0y}/p_{0x}) = \pi/2 - \theta_0$.

To find the dispersion in the neighborhood of the point $p_x = p_y = 0$, we employ the perturbation theory. Treating $p_x$ and $p_y$ as small perturbations, we expand all quantities to the first order $E = E^{(0)} +  E^{(1)} + ...;~A = A^{(0)} + A^{(1)} + ...;~B = B^{(0)} + B^{(1)} + ...$. We obtain a cross-like solution describing the nonlinear Dirac cone: 
\begin{equation} \label{eq:E0E1}
E = E_3^{(0)} + E^{(1)} =  \frac{ \Gamma I_0}{2} \pm \frac{ \sqrt{2} J_1 \sqrt{ p_x^2 + p_y^2}}{\sqrt{1 - \frac{4 \left( \Delta - 4 J_2\right)^2}{I_0^2 \Gamma^2}}}.
\end{equation}
Thus, at $\Gamma^2 I_0^2>4 \left( \Delta - 4 J_2\right)^2$ one of the dispersion curves develops a loop. 

The intensities of two components nearby the cross point are corrected as follows 
\begin{gather} \label{eq:ABp}
|A|^2  = \dfrac{I_0}{2} - \dfrac{\Delta - 4 J_2}{\Gamma} \left( 1 \mp \frac{ 2 \sqrt{2} J_1 \sqrt{ p_x^2 + p_y^2}}{\sqrt{I_0^2 \Gamma^2  - {4 (\Delta - 4 J_2)^2} }}  \right),~|B|^2  = \dfrac{I_0}{2} + \dfrac{\Delta - 4 J_2}{\Gamma} \left( 1 \mp \frac{ 2 \sqrt{2} J_1 \sqrt{ p_x^2 + p_y^2}}{\sqrt{I_0^2 \Gamma^2  - {4 (\Delta - 4 J_2)^2} }}  \right).
\end{gather}
Substituting amplitudes~\eqref{eq:ABp} and energy~\eqref{eq:E0E1} into the system~\eqref{eq:e_vect}, we may introduce the local effective Hamiltonian as:
\begin{equation} 
\label{eq:per_dirac_1st_p}
\begin{pmatrix} 
\pm 2 \sqrt{2} J_1\frac{ (\Delta - 4 J_2)\sqrt{ p_x^2 + p_y^2}}{\sqrt{I_0^2 \Gamma^2  - {4 (\Delta - 4 J_2)^2} }} & - \sqrt{2} J_1(p_y + i p_x)  \\
 -\sqrt{2} J_1(p_y - i p_x)  & \mp 2 \sqrt{2} J_1\frac{  (\Delta - 4 J_2) \sqrt{ p_x^2 + p_y^2}}{\sqrt{I_0^2 \Gamma^2  - {4 (\Delta - 4 J_2)^2} }}
\end{pmatrix}
\begin{pmatrix} \psi_1  \\ \psi_2 \end{pmatrix} =  E^{(1)} \begin{pmatrix} \psi_1  \\  \psi_2 \end{pmatrix} = \pm \frac{ \sqrt{2} J_1 \sqrt{ p_x^2 + p_y^2}}{\sqrt{1 - \frac{4 \left( \Delta - 4 J_2\right)^2}{I_0^2 \Gamma^2}}}\begin{pmatrix} \psi_1  \\  \psi_2 \end{pmatrix} \:.
\end{equation}

Next, we derive the exact implicit expression for the nonlinear dispersion $E(p_x,p_y)$. To simplify our derivations, we set $p_x=0$ and rewrite the system in the form: 
\begin{equation} 
\begin{pmatrix} 
{E_n - M_n } &  J_1 \sqrt{2} p_y \\
 J_1 \sqrt{2} p_y  & {E_n + M_n}
\end{pmatrix}
\begin{pmatrix} A \\  B \end{pmatrix} =0 \:,
\end{equation}
denoting $E_n = E- \Gamma I_0 /2$, $M_n   = \Delta - 4 J_2 +  \dfrac{\Gamma I_0 \left( \Delta - 4 J_2 +  J_2 p_y^2 \right ) }{2 (E - \Gamma I_0)} + J_2 p_y^2 $. 
The nonlinear dispersion is then given by
\begin{equation} 
\label{eq:non_dis_imp}
E_n^2 = 2 J_1^2 p_y^2 + M_n^2 (p_y^2).  
\end{equation}
The eigenvectors' intensities on the two sublattices satisfy 
\begin{align}
\label{eq:amp_k_dep_mass}
|A|^2  = \dfrac{I_0}{2}+\dfrac{\left( \Delta - 4 J_2 \right) I_0}{2( -\Gamma I_0  + E)} + \dfrac{J_2 p_y^2 I_0}{2( - \Gamma I_0  + E) },\\
|B|^2  = \dfrac{I_0}{2}-\dfrac{\left( \Delta - 4 J_2 \right) I_0}{2( -\Gamma I_0  + E)} -\dfrac{J_2 p_y^2 I_0}{2( - \Gamma I_0  + E) }.
\end{align}
The implicit relation~\eqref{eq:non_dis_imp} can be posed as
\begin{equation}
((E - \Gamma I_0/2)^2 - 2 J_1^2 p_y^2) \left(E- \Gamma I_0  \right)^2 =\left(\Delta - 4 J_2 + J_2 p_y^2 \right) ^2\left( E - \dfrac{ \Gamma I_0}{2}\right)^2.
\end{equation}
Note, this dispersion relation supports the existence of 2 more loops in addition to the loop at the point $p_y=p_{\text{I}}=0$, described above. This bifurcation occurs at $d_z (\bm{p})=0$:
\begin{equation}
p_y=\pm p_{\text{II}}=\pm\sqrt{\dfrac{4 J_2 - \Delta}{J_2}},
\end{equation}
in the nontrivial phase only, $|\Delta|<4J_2$.
The energies at the points $p_y = \pm p_{\text{II}}$ are: 
\begin{gather}
\label{eq:Eu0_BHZ}
E^{(0)\text{II}}_{3}=\Gamma I_0,\\
E^{(0)\text{II}}_{2,1}=\Gamma I_0/2 \pm \sqrt{ \dfrac{2 J_1^2 (4 J_2 - \Delta)}{J_2}}.
\end{gather}
Specifically, the energy $E^{(0){\text{II}}}_3$ corresponds to two additional cross points, 
which appear only in the nontrivial case with the eigenvectors
\begin{equation}
\left( \!{\begin{array}{*{20}{c}}
{ A^{(0)\text{II}}} \\
{ B^{(0)\text{II}}}
 \end{array}} \!  \right) =
\left( \!{\begin{array}{*{20}{c}}
{ e^{i \varphi} \sqrt{\dfrac{ I_0}{2} + \sqrt{\dfrac{ I_0^2}{4} - 2 J_{1}^2 (p_y^{\text{II}})^2/\Gamma^2 }}} \\
{  \pm \sqrt{\dfrac{ I_0}{2} - \sqrt{\dfrac{ I_0^2}{4} - 2 J_{1}^2 (p_y^{\text{II}})^2/\Gamma^2 }}}
\end{array}} \!  \right). 
\end{equation}
The additional crosses appear at the intensities higher $\Gamma I_0 /2 =  \pm \sqrt{ \dfrac{2 J_1^2 (4 J_2 - \Delta)}{J_2}}$ (the sign is chosen depending on sign of the nonlinearity $\Gamma$), which is defined by the degeneracy of the cross point and one of the bands at $p_y=\pm p_{\text{II}}$. 

\subsection{Modulation instability} 

To examine linear stability of the nonlinear Bloch modes, we introduce small complex-valued perturbations to the amplitudes: $A = A_0+ \delta a$, $B = B_0 + \delta b$ and look for the solution in the form: 
\begin{equation}
\left( \!{\begin{array}{*{20}{c}}
{ \psi_1 } \\
{ \psi_2}
 \end{array}} \!  \right) =
 \left( \!{\begin{array}{*{20}{c}}
{ A_0 + \delta a} \\
{ B_0 + \delta b}
\end{array}} \!  \right)  e^{- i E t + i p_x x + i p_y y}.
\end{equation}
The equations for deviations $\delta a,~\delta b$ can be recast as
\begin{equation}
\label{eq:main_for_a_a_conj_b_b_conj}
i\dfrac{\partial}{\partial t}  \begin{pmatrix} \delta  a  \\  \delta  b \\ \delta  a^* \\ \delta  b^* \end{pmatrix}  = {\hat{ L}} \begin{pmatrix} \delta  a  \\  \delta  b \\ \delta  a^* \\ \delta  b^* \end{pmatrix} \:,
\end{equation}
where operator ${\hat{ L}}$ is the $ 4 \times 4 $ matrix
\begin{equation}
{\hat{ L}} =  \begin{pmatrix} 
\hat{H}_{D} (\partial_x,\partial_y) + {H}_{D} (p_x,p_y)- E \hat{I} + 2 \Gamma  \begin{pmatrix} |A_0|^2 & 0 \\ 0  & |B_0|^2 \end{pmatrix}   & \Gamma \begin{pmatrix} A_0^2 & 0 \\ 0  & B_0^2 \end{pmatrix} \\
- \Gamma  \begin{pmatrix} A^{*2}_0 & 0 \\ 0  & B_0^{*2} \end{pmatrix} & -\hat{H}^{*}_{D}(\partial_x,\partial_y)  - {H}^{*}_{D} (p_x,p_y) + E \hat{I} - 2 \Gamma \begin{pmatrix} |A_0|^2 & 0 \\ 0  & |B_0|^2 \end{pmatrix}
\end{pmatrix}  ,
\end{equation} 
where \begin{equation}
{H}_{D} (p_x,p_y) = 
\begin{pmatrix} 
J_2 \left(p_x^2+p_y^2 \right)&-J_{1}\sqrt{2}\left( p_y + i p_x \right)\\
-J_{1}\sqrt{2}\left( p_y - i p_x \right)& - J_2 \left(p_x^2+p_y^2 \right)\\
\end{pmatrix},
\end{equation}
\begin{equation}
\hat{H}_{D} (\partial_x,\partial_y) = 
\begin{pmatrix} 
\Delta - 4 J_2 -J_2 \left(\partial_x^2+\partial_y^2 \right)&J_{1}\sqrt{2}\left( i \partial_y - \partial_x \right)\\
J_{1}\sqrt{2}\left( i \partial_y + \partial_x \right)&-\Delta + 4 J_2 J_2 \left(\partial_x^2+\partial_y^2 \right)\\
\end{pmatrix}.
\end{equation}

To study modulational instability, we take  $[\delta a; \delta a^*; \delta b; \delta b^*] = [\bar{C}_1; \bar{C}_2; \bar{C}_3;\bar{C}_4] e^{ - i \lambda t + i\kappa_x x + i\kappa_y y} =  \bar{{\boldsymbol{C}}} e^{ - i \lambda t + i\kappa_x x + i\kappa_y y}  $ and set $\kappa_x=0$ to simplify further considerations. Eq.~\eqref{eq:main_for_a_a_conj_b_b_conj} leads to the system of equations for amplitudes $\bar{\boldsymbol{C}}$: $\left( \hat L - \lambda \hat I \right)\bar{\boldsymbol{C}} = 0 $. The positive imaginary part of $\lambda$, found from $\det \left( \hat L - \lambda \hat I \right)  =0$, indicates instability. 

For the cross point at $p_x=p_y=0$, existing at the intensities $ \Gamma I_0 >\pm 2 \left( \Delta - 4 J_2 \right)$ at the energy $E^{(0)} =   \frac { \Gamma I_0}{2}$, with the amplitudes $|A_0|^2 = \dfrac{I_0}{2}  -  \dfrac{ \Delta-4 J_{2}  }{\Gamma} $, $|B_0|^2 = \dfrac{I_0}{2}  + \dfrac{ \Delta-4 J_{2} }{\Gamma}$, we find the energy detuning $\lambda$ along the straight lines $I_0 \Gamma + C = - 2(\Delta - 4 J_2)$ in the parameter plane $(\Gamma, \Delta)$:

\begin{equation}
\label{eq:lam_cr_with_phi}
\lambda=\pm \sqrt{\pm \dfrac{\sqrt{e^{-2 i \varphi} \kappa_y^2 \left(C^2 \left(-1+e^{2 i \varphi}\right)^2 J_1^2+2 C \left(-1+e^{2 i \varphi}\right)^2 \Gamma I_0 J_1^2 \right)+2\Gamma^2 I_0^2 J_2^2 \kappa_y^4}}{\sqrt{2}}+ C J_2 \kappa_y^2+ \Gamma I_0 J_2 \kappa_y^2+2 J_1^2 \kappa_y^2+ J_2^2 \kappa_y^4}\:.
\end{equation}

We analyse Eq.~\eqref{eq:lam_cr_with_phi} for $C=0$ at the line $I_0 \Gamma  = - 2(\Delta - 4 J_2)$, which is the negatively inclined existence boundary of the cross solution:
\begin{gather} \label{}
\lambda_{1,2}=\pm\sqrt{J_2^2 \kappa_y^4+2\kappa_y^2 J_1^2}, \label{eq:lambda_cross_1}\\
\lambda_{3,4}=\pm\sqrt{- 4 (\Delta - 4 J_2) J_2 \kappa_y^2 +J_2^2 \kappa_y^4+ 2 J_1^2 \kappa_y^2}. \label{eq:lambda_cross_2}
\end{gather}
The imaginary part $\Im (\lambda_{1,2})$ is zero for all values of the wave number $\kappa_y$, therefore, $\lambda_{1,2}$ do not show any instability. The area of the stability can be determined from $\lambda_{3,4}$: it is a purely real quantity for $\Gamma > -\dfrac{J_1^2}{I_0 J_2}$ or equivalently $2J_1^2 \geq 4 \left( \Delta - 4 J_2 \right) J_2$. In the nontrivial case, since  $J_2 \left( \Delta - 4 J_2\right)<0$, we conclude that $2J_1^2 \geq 4 \left( \Delta - 4 J_2 \right) J_2$ for any $J_2, \Delta$. Therefore, the cross point is stable. But in the trivial case, the area of parameters $J_2, J_1$ exists, for which $\Im (\lambda_{3,4})>0$, and the cross point becomes unstable. The boundary value of the detuning in the trivial phase is
\begin{equation} 
\Delta_c = 4J_2 + \dfrac{J_1^2}{2 J_2}.
\end{equation}
Note, for the given intensity, $- \Gamma I_0/2=\Delta-4 J_{2} $, the upper branch and the point of the cross are degenerate. Hence, the line of stability $I_0 \Gamma  = - 2(\Delta - 4 J_2)$ appears in Fig.~1 in the main text for the upper branch at $\Delta < \Delta_c$. For $\Delta > \Delta_c$, we analytically obtain the maximum growth rate $\max_{\kappa_y} \Im [\lambda_{3,4}]$ achieved at the wavenumber $\kappa_y^{\text{max}}$:
\begin{gather}
\label{eq:inc_an}
\max_{\kappa_y} \Im [\lambda_{3,4}] = \frac{|J_1^2 + \Gamma I_0 J_2|}{ |J_2|};\\
\kappa_y^{\text{max}} = \pm \sqrt{\frac{ |\Gamma I_0 J_2 +J_1^2|}{J_2^2}}.
\end{gather}

Equation~\eqref{eq:lam_cr_with_phi} at the other boundary of the existence of the cross solution (with $C= - 2 \Gamma I_0$) takes the form:
\begin{equation}
\lambda_{1,2,3,4}=\pm \sqrt{\pm \kappa_y^2  \Gamma I_0 J_2-\Gamma I_0 J_2 \kappa_y^2+2 J_1^2 \kappa_y^2+ J_2^2 \kappa_y^4},
\end{equation} 
from which we obtain the area of stability $\Gamma < \dfrac{J_1^2}{I_0 J_2}$. 

Let us consider Eq.~\eqref{eq:lam_cr_with_phi} for the case $\varphi = \pi n, n \in \mathbb{Z}$:
\begin{gather}
\lambda_{1,2} = \pm \sqrt{J_2^2\kappa_y^4  + C J_2 \kappa_y^2 + 2 J_1^2 \kappa_y^2},\\
 \lambda_{3,4}= \pm \sqrt{J_2^2\kappa_y^4  + C J_2 \kappa_y^2 + 2 J_2 \Gamma I_0 \kappa_y^2+ 2 J_1^2 \kappa_y^2}.
\end{gather}
The boundaries of the cross stability are located on lines with
$C= - \dfrac{2 J_1^2}{J_2} - 2 \Gamma I_0$ and $C=- \dfrac{2 J_1^2}{J_2}$.
These are the straight lines $I_0 \Gamma = \pm 2 (\Delta - 4 J_2 - \frac{J_1^2}{J_2})$. 

\begin{figure}[hb] 
\center{\includegraphics[width=1\linewidth]{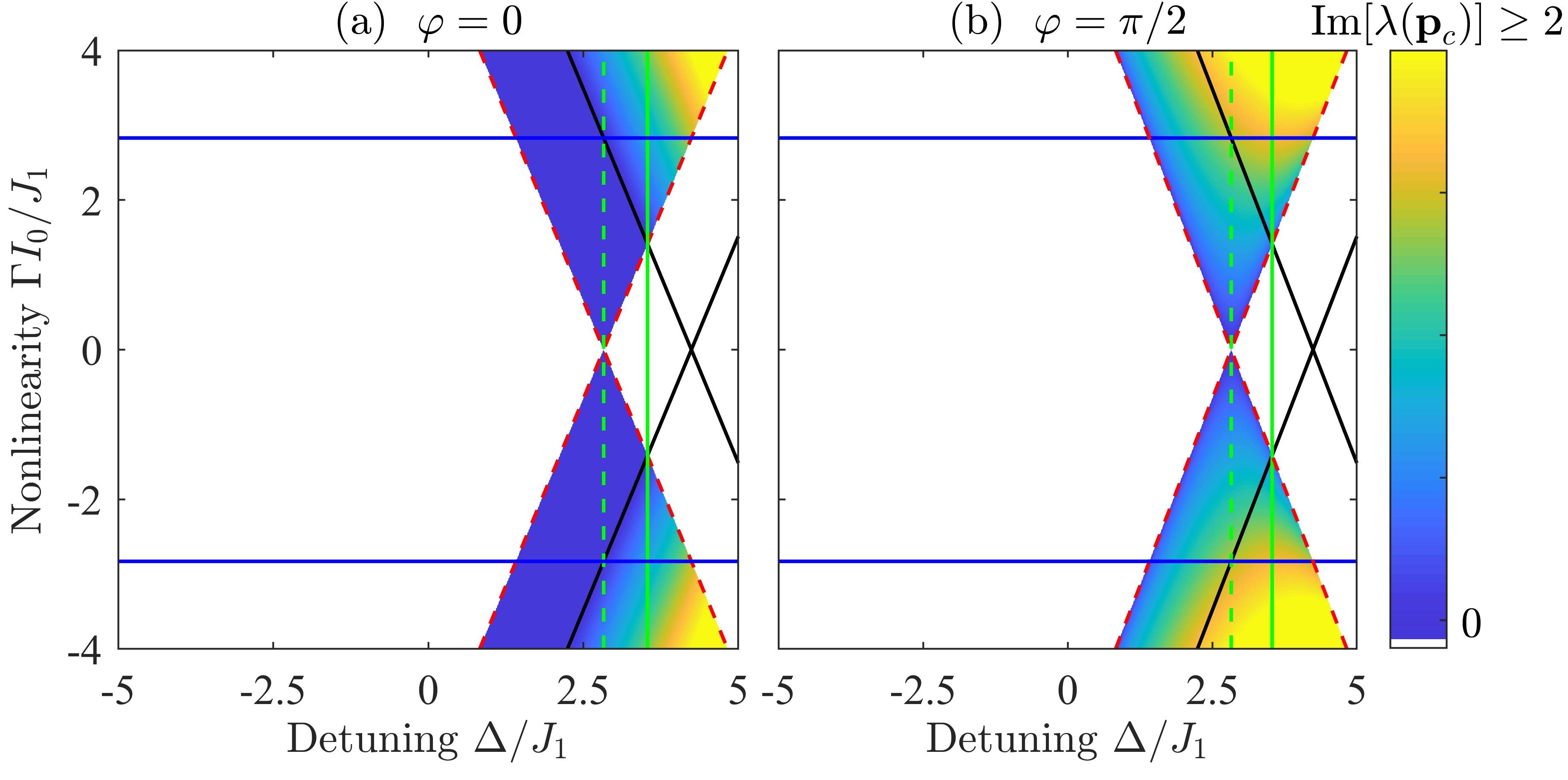}} 
\caption{ The  maximum increment value $\max_{\lambda} [\Im (\lambda)] $ color-coded in the plane of parameters $\Delta/J_1,~\Gamma I_0/J_1$ for the cross point $E = \Gamma I_0/2$ at $p=0$. 
Parameters are $J_2=J_1/\sqrt{2}$, 
(a) $\varphi =0$, (b) $\varphi = \pi/2$. 
Red dashed lines $\Gamma = \pm 2\left( 4 J_2 - \Delta\right)/I_0$ highlight the boundaries of the existence of the cross solution. 
On these lines, the cross point is stable in the nontrivial domain, $|\Delta|<2 \sqrt{2}$, whose upper boundary is marked with the green dashed line. 
In the trivial domain, the cross point is unstable at detunings larger $\Delta = {J_1^2}/{2 J_2} + 4 J_2 $ marked with a solid green line. 
The boundaries of the cross point stability for $\varphi=0$ are black straight lines $I_0 \Gamma = \pm 2 (\Delta - 4 J_2 - {J_1^2}/{J_2})$. At $\varphi = 0$,  the intersection point of the black lines with the boundary of the trivial phase at $\Gamma = \pm {2 J_1^2}/{(J_2 I_0)}$ (straight blue lines) defines the intensity, for which, by changing $\Delta$, we can distinguish the trivial phase from the nontrivial one by observing a transition from stability to instability.  
\label{fig:figS1}
}
\end{figure}

The color maps of the maximum increment value $\max_{\lambda} [\Im (\lambda)] $ in the parameter space for the cross solution $E = \Gamma I_0/2$ are plotted in Fig.~\ref{fig:figS1} by using Eq.~\eqref{eq:lam_cr_with_phi}. On our notations, linear stability of perturbation in $y$ direction depends on the spinor phase angle $\varphi$. 
Making a generalization about this feature, we note that stability is conditional on the mutual orientation $\Delta \theta = \theta_p - \theta_0 $ of the symmetry-broken solution with ${\bm p}_0$ and perturbation with ${\bm p}_p$. In the nontrivial phase, it remains stable at $\Delta \theta = 0$ ($\varphi = 0$, $\theta_p = \pi/2$, $\theta_0 = \pi/2$) and exhibits the maximum transverse modulational instability at $\Delta \theta = \pi/2$ ($\varphi = \pi/2$, $\theta_p = \pi/2$, $\theta_0 = 0$).

Next, we describe modulational instability of the the background state $p_x = p_y =0$ with a uniform intensity $\Gamma I_0 < \pm 2 \left( \Delta - 4 J_2 \right) $. 
For definiteness, we consider the nonlinear mode from the band with energy $E = \left( \Delta-4 J_{2} \right)  + \Gamma I_0 $, and spinor components $B^{(0)} =0$,  $|A^{(0)}|^2 =I_0$ [see Eq.~\eqref{eq:Eu0}]. The solutions of equation $\det \left( \hat L - \lambda \hat I \right)  =0$ along the lines $I_0 \Gamma + C = - 2(\Delta - 4 J_2)$ recast as 
\begin{gather}
\label{eq:lam_line}
\lambda_{1,2}=\pm\sqrt{-F+D}\:,\\
\lambda_{3,4}=\pm\sqrt{F+D}\:,
\end{gather}
being expressed through the auxiliary functions
\begin{gather}
 F=\sqrt{C^4/4+(C+\Gamma I_0)^2 J_2^2 \kappa_y ^4 + \kappa_y^2(-J_2 C^2 (C+\Gamma I_0)+4 J_1 C \Gamma I_0+2 C^2 J_1^2)}\:,\\
 D=J_2^2 \kappa_y^4+\kappa_y^2(J_2 I_0 \Gamma+2 J_1^2-C J_2)+C^2/2.
\end{gather}
At $C=0$ we recover \eqref{eq:lambda_cross_1} and \eqref{eq:lambda_cross_2} for the cusp bifurcation point. Varying parameter $C$, one can reproduce the numerically-obtained map of instability shown in Fig.~1(c) of the main text. In the linear limit $\Gamma I_0 \rightarrow 0$, we obtain the spectrum  
\begin{equation}
\lambda = \pm \left[- m_\text{eff} \pm \sqrt{ (m_\text{eff} + J_2 \kappa_y^2)^2 + 2 J_1^2 \kappa_y^2 } \right]\:, 
\end{equation}
where $m_\text{eff}  = \Delta - 4 J_2$.
Assuming $|\lambda| \sim I_0 \sim \kappa_y^2 \sim \mu \ll 1$, where $\mu$ is the smallness parameter, from the determinant calculated to quadratic accuracy $\sim \mu^2$, 
we get 
\begin{equation}
\lambda^2 = \left(\Gamma I_0 + J_2 \kappa_y^2 + \dfrac{ J_1^2 \kappa_y^2}{m_{\text{eff}}} \right)^2 - \Gamma^2 I_0^2 =  \kappa_y^2 \left( J_2 + \dfrac{J_1^2}{m_{\text{eff}}}\right) \left[ 2\Gamma I_0 + \left(J_2 + \dfrac{J_1^2}{m_{\text{eff}}}  \right) \kappa_y^2 \right] \:. 
\end{equation}
In the trivial phase $m_{\text{eff}}>0$, similar to the nonlinear Schr\"{o}dinger equation, instability only occurs for the self-focusing nonlinearity ($\Gamma < 0$), and the maximum increment of instability 
is achieved at $\kappa_y^2 = - {\Gamma I_0} / \left(J_2 +  {J_1^2}/{m_{\text{eff}}} \right)$
with purely imaginary $\lambda = \pm i |\Gamma| I_0 $. On the other hand, in the nontrivial phase $m_{\mathrm{eff}} < 0$, for a fixed $\Gamma$ the state can either be stable or unstable, depending on $|m_{\mathrm{eff}}|$.

\begin{figure} [h]
    \centering
    \includegraphics[width=0.85\linewidth]{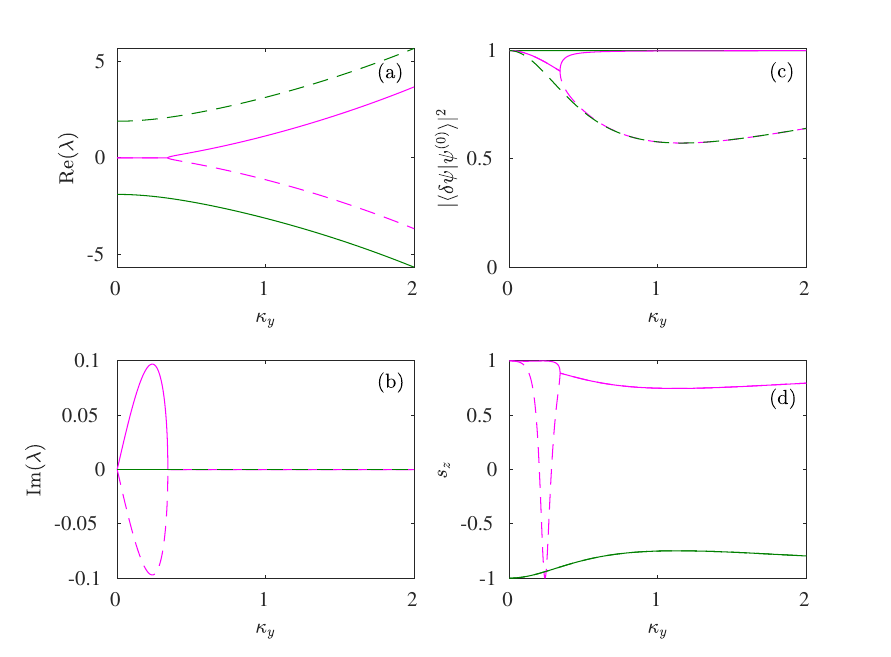}
    \caption{(a) Real and (b) imaginary parts of linear perturbation eigenvalues $\lambda$, (c) squared magnitude of the overlap between the perturbation modes and linear Bloch waves, (d) spin $s_z$ projection of perturbation eigenvectors for the modulational instability in the weakly nonlinear regime, focusing nonlinearity $\Gamma = -1$, $I_0 = 0.1$, $m_{\text{eff}}=1$ (trivial).}
    \label{fig:MI_poln_trivial}
\end{figure}

\begin{figure} [h]
    \centering
    \includegraphics[width=0.85\linewidth]{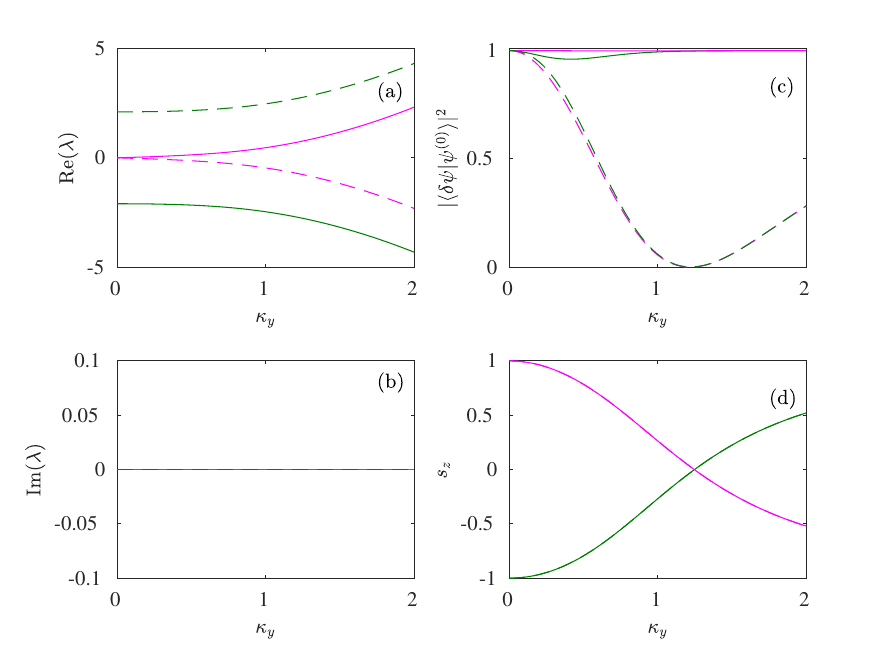}
    \caption{(a) Real and (b) imaginary parts of linear perturbation eigenvalues $\lambda$, (c) squared magnitude of the overlap between the perturbation modes and linear Bloch waves, (d) spin $s_z$ projection of perturbation eigenvectors in the weakly nonlinear regime, focusing nonlinearity $\Gamma = -1$, $I_0 = 0.1$, $m_{\text{eff}}=-1$ (nontrivial).}
    \label{fig:MI_poln_nontrivial}
\end{figure}

Figs.~\ref{fig:MI_poln_trivial} and~\ref{fig:MI_poln_nontrivial} illustrate dispersion and polarization of perturbation modes as a function of $\kappa_y$ in the trivial and nontrivial phases in the weakly nonlinear regime. We characterize the perturbation modes' polarization by computing their time-averaged overlap with the linear Bloch waves $|\braket{\delta \psi| \psi^{(0)}}|^2$ and their sublattice imbalance $s_z =|\delta a|^2 - |\delta b|^2 = \braket{\sigma_z}$. In the trivial phase the perturbation modes maintain a moderate to large overlap with the linear Bloch waves, and $|s_z|$ remains large, indicating the perturbation modes are preferentially localized to a single sublattice. On the other hand, in the nontrivial phase $|s_z|$ rotates to the opposite pole of the Bloch sphere at large $\kappa_y$, reducing the effective strength of the nonlinearity-induced wave mixing.

\section{\large{Modulational instability dynamics and soliton formation}}

In this section we present additional details of the modulational instability dynamics summarised in Fig.~3 of the main text. We further characterize the dynamics by providing snapshots of the field intensity profiles in real and Fourier space at various times, as well as computing the linear and nonlinear parts of the conserved total energy $H = H_L + H_{NL}$,
\begin{equation}
    H_L(t) = \int d\bs{k} \braket{\psi(\bs{k},t)|\hat{H}_L(\bs{k})|\psi(\bs{k},t)}, \quad H_{NL}(t) = \int d\bs{r} \braket{\psi(\bs{r},t)|\hat{H}_{NL}(\bs{r})|\psi(\bs{r},t)}.
\end{equation}
$H_L$ provides a measure of which linear modes are excited by the wave field, while $H_{NL}$ is sensitive to the field's localization.

\begin{figure}
    \centering
    \includegraphics[width=\linewidth]{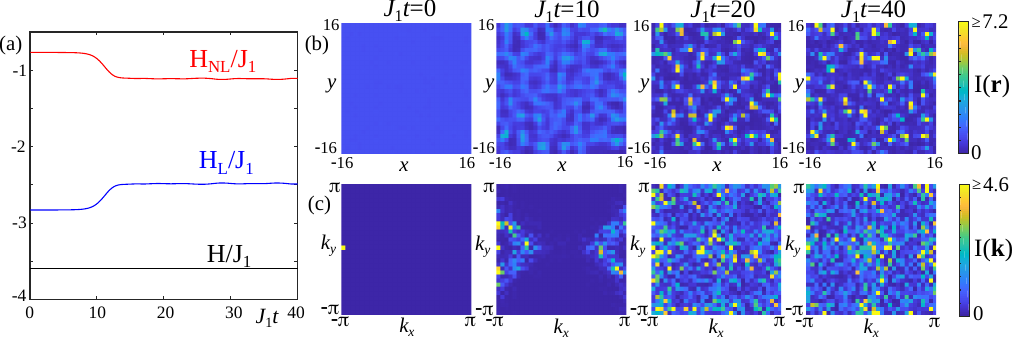}
    \caption{Dynamics of the modulational instability in the focusing regime, $\Delta = 0, \Gamma = -1.25$. (a) Time evolution of the linear $H_L$, nonlinear $H_{NL}$, and total $H = H_L + H_{NL}$ energies of the weakly-perturbed nonlinear Bloch wave. (b) Snapshots of the real space intensity $I(\bs{r}) = |\psi_a(\bs{r})|^2 + |\psi_b(\bs{r})|^2$ at different propagation times. (c) Snapshots of the Fourier space intensity $I(\bs{k}) = |\psi_a(\bs{k})|^2 + |\psi_b(\bs{k})|^2$ at different propagation times. In (b,c) the maximum of the colour scale is chosen to be 50\% of the peak intensity at $J_1 t = 40$ in order to enhance the visibility of the field at the shorter times. Other parameters are the same as in Fig. 3 of the main text.}
    \label{fig:focusing_dynamics}
\end{figure}

Fig.~\ref{fig:focusing_dynamics} illustrates the dynamics of a perturbed nonlinear Bloch wave in the focusing instability regime ($\Delta = 0$, $\Gamma = -1.25$). For this choice of $\Delta$ the nonlinear Bloch wave lies at the lower edge of the lower band. At short times there is a clear amplification of the linearly unstable perturbation modes, which have wavevectors close to $\bs{k}_0$. As the instability continues to develop, a significant amount of energy is transferred to other wavevectors throughout the entire Brillouin zone, resulting in an increase of $H_L$. At the same time, strongly-localized soliton-like structures develop in real space, increasing $|\hat{H}_{NL}|$, as required for conservation of $H$. Thus, under the focusing nonlinearity the entire band becomes excited.

\begin{figure}
    \centering
    \includegraphics[width=\linewidth]{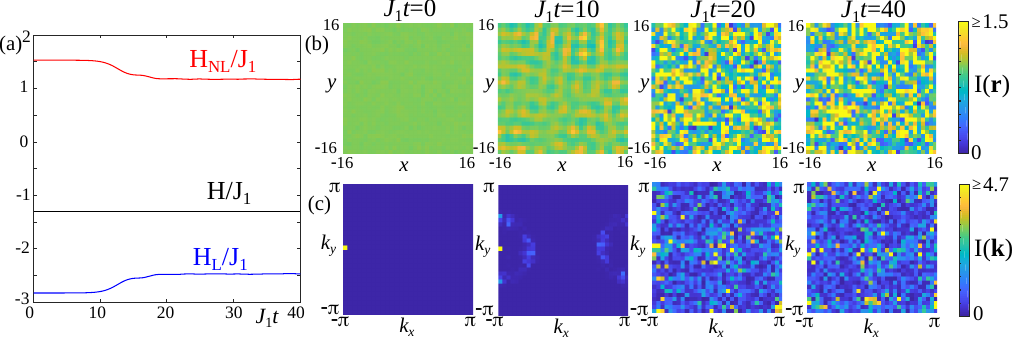}
    \caption{Dynamics of the modulational instability in the defocusing regime, $\Delta = 0, \Gamma = 2.5$. (a) Time evolution of the linear $H_L$, nonlinear $H_{NL}$, and total $H = H_L + H_{NL}$ energies of the weakly-perturbed nonlinear Bloch wave. (b) Snapshots of the real space intensity $I(\bs{r}) = |\psi_a(\bs{r})|^2 + |\psi_b(\bs{r})|^2$ at different propagation times. (c) Snapshots of the Fourier space intensity $I(\bs{k}) = |\psi_a(\bs{k})|^2 + |\psi_b(\bs{k})|^2$ at different propagation times. In (b,c) the maximum of the colour scale is chosen to be 50\% of the peak intensity at $J_1 t = 40$ in order to enhance the visibility of the field at the shorter times. Other parameters are the same as in Fig. 3 of the main text.}
    \label{fig:defocusing_dynamics}
\end{figure}

Next, we show in Fig.~\ref{fig:defocusing_dynamics} the dynamics in the defocusing instability regime ($\Delta = 0$, $\Gamma = 2.5$). For short times we see a similar amplification of wavevectors close to $\bs{k}_0$. At longer times, there is a significant transfer of energy throughout the entire Brillouin zone, increasing $H_L$. However, in this case the dynamics decrease $|H_{NL}|$, indicating the field delocalizes in real space, distributing energy between both of the sublattices, and does not form any soliton-like structures. Interestingly, despite the qualitative differences between the real space field distribution in the focusing and defocusing cases, both exhibit a spreading of energy in Fourier space throughout the entire lower band, generating a large purity gap and allowing measurement of the band's Chern number.

\begin{figure}
    \centering
    \includegraphics[width=\linewidth]{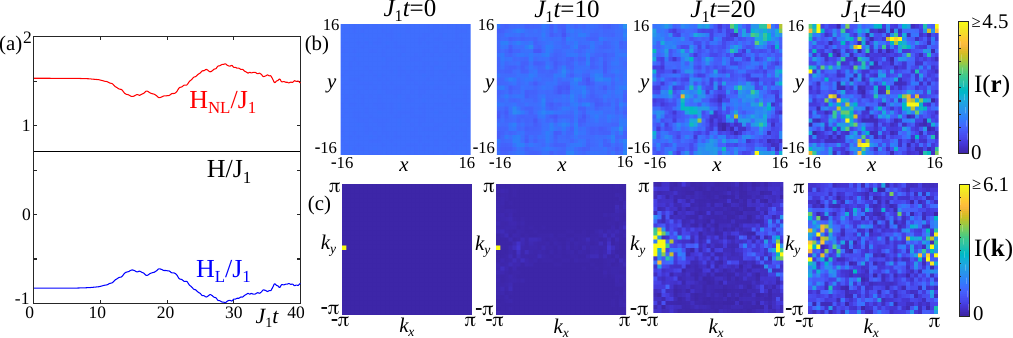}
    \caption{Dynamics of the modulational instability in the oscillatory instability regime, $\Delta = 2, \Gamma = 2.5$. (a) Time evolution of the linear $H_L$, nonlinear $H_{NL}$, and total $H = H_L + H_{NL}$ energies of the weakly-perturbed nonlinear Bloch wave. (b) Snapshots of the real space intensity $I(\bs{r}) = |\psi_a(\bs{r})|^2 + |\psi_b(\bs{r})|^2$ at different propagation times. (c) Snapshots of the Fourier space intensity $I(\bs{k}) = |\psi_a(\bs{k})|^2 + |\psi_b(\bs{k})|^2$ at different propagation times. In (b,c) the maximum of the colour scale is chosen to be 50\% of the peak intensity at $J_1 t = 40$ in order to enhance the visibility of the field at the shorter times. Other parameters are the same as in Fig. 3 of the main text.}
    \label{fig:oscillatory_dynamics}
\end{figure}

Finally, Fig.~\ref{fig:oscillatory_dynamics} shows the dynamics in the oscillatory instability regime ($\Delta = 2$, $\Gamma = 2.5$). Again, at short times there is an amplification of wavevectors close to $\bs{k}_0$, before the entire Brillouin zone becomes populated at long times. For this $\Delta$ the initial Bloch wave lies at the upper edge of the lower band. Thus, the initial increase in $H_L$ indicates a significant transfer of energy to the upper band. As the instability progresses, however, the spreading of energy within the lower band begins to dominate, leading to a decrease of $H_L$. In this case neither $H_{L}$ nor $H_{NL}$ converge to a stationary value and no stationary soliton-like structures are visible in the real space intensity profile.

\begin{figure}
    \centering
    \includegraphics[width=0.5\linewidth]{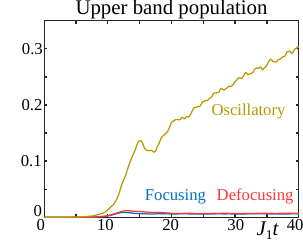}
    \caption{Dynamics of the upper band populations $\int d\bs{k} |\braket{u_+(\bs{k})|\psi(\bs{k},t)}|^2$ in the three regimes shown in Fig. 3 of the main text.}
    \label{fig:mixing_dynamics}
\end{figure}

As a further check, we also quantified the strength of interband mixing by calculating the time-dependent population of the upper band, $\int d{\bs{k}} |\braket{u_+(\bs{k}|\psi(\bs{k},t)}|^2$, with results shown in Fig.~\ref{fig:mixing_dynamics}. In the focusing and defocusing instability examples, the interband mixing is negligible ($\approx 1\%$ of the total field energy), whereas in the case of oscillatory instability there is a significant transfer of energy into the upper band, which persists for long times. This strong energy transfer occurs due to the nonlinearity-induced band inversion, which enables resonant inter-band wave mixing.

\begin{thebibliography}{99}

\bibitem{Topo_review}
T. Ozawa et al., {\it Topological photonics}, Rev. Mod. Phys. {\bf 91}, 015006 (2019).

\bibitem{NonlinearTopo_review}
D. Smirnova, D. Leykam, Y. D. Chong, and Y. Kivshar, {\it Nonlinear topological photonics}, Appl. Phys. Rev. {\bf 7}, 021306 (2020).

\bibitem{review2}
A. Saxena, P. G. Kevrekidis, and J. Cuevas-Marave, {\it Nonlinearity and Topology}, Nonlinear Systems and Complexity {\bf 32}, 25 (2020).

\bibitem{polariton_topo}
S. Klembt et al., {\it Exciton-polariton topological insulator}, Nature {\bf 562}, 552 (2018).


\bibitem{Baboux2018}
F. Baboux et al., {\it Unstable and stable regimes of polariton condensation}, Optica {\bf 5}, 1163 (2018).

\bibitem{Mukherjee_arxiv}
S. Mukherjee and M. C. Rechtsman, {\it Observation of Floquet solitons in a topological bandgap}, Science {\bf 368}, 856 (2020).

\bibitem{Smirnova2019}
D. A. Smirnova et al., {\it Third-harmonic generation in photonic topological metasurfaces}, Phys. Rev. Lett. {\bf 123}, 103901 (2019).

\bibitem{topological_source}
S. Mittal, E. A. Goldschmidt, and M. Hafezi. {\it A topological source of quantum light}, Nature {\bf 561}, 502 (2018).

\bibitem{Ablowitz2014}
M. J. Ablowitz, C. W. Curtis, and Y.-P. Ma, {\it Linear and nonlinear traveling edge waves in optical honeycomb lattices}, Phys. Rev. A {\bf 90}, 023813 (2014).

\bibitem{Lumer2016}
Y. Lumer, M. C. Rechtsman, Y. Plotnik, and M. Segev, {\it Instability of bosonic topological edge states in the presence of interactions}, Phys. Rev. A {\bf 94}, 021801(R) (2016).

\bibitem{Kartashov2016}
Y. V. Kartashov and D. V. Skryabin, {\it Modulational instability and solitary waves in polariton topological insulators}, Optica {\bf 3}, 1228 (2016).

\bibitem{Leykam2016}
D. Leykam and Y. D. Chong, {\it Edge Solitons in Nonlinear Photonic Topological Insulators}, Phys. Rev. Lett. {\bf 117}, 143901 (2016).


\bibitem{Lumer2013}
Y. Lumer, Y. Plotnik, M. C. Rechtsman, and M. Segev, {\it Self-Localized States in Photonic Topological Insulators}, Phys. Rev. Lett. {\bf 111}, 243905 (2013).

\bibitem{Poddubny2018}
A. N. Poddubny and D. A. Smirnova, {\it Ring Dirac solitons in nonlinear topological systems}, Phys. Rev. A {\bf 98}, 013827 (2018).

\bibitem{Marzuola_arxiv}
J. L. Marzuola, M. Rechtsman, B. Osting, and M. Bandres, {\it Bulk soliton dynamics in bosonic topological insulators}, arXiv:1904.10312.


\bibitem{Smirnova2019b}
D. A. Smirnova, L. A. Smirnov, D. Leykam, and Y. S. Kivshar, {\it Topological edge states and gap solitons in the nonlinear Dirac model}, Laser Photon. Rev. {\bf 13}, 1900223 (2019).

\bibitem{MI_review}
V. E. Zakharov and L. A. Ostrovsky, {\it Modulation instability: The beginning}, Physica D {\bf 238}, 540 (2009).

\bibitem{NL_book}
R. W. Boyd, {\it Nonlinear Optics}, Academic Press, New York (2008).

\bibitem{Kivshar1992}
Y. S. Kivshar and M. Peyrard, {\it Modulational instabilities in discrete lattices}, Phys. Rev. A {\bf 46}, 3198 (1992).


\bibitem{Engelhardt2015}
\revision{G. Engelhardt and T. Brandes, {\it Topological Bogoliubov excitations in inversion-symmetric systems of interacting bosons}, Phys. Rev. A {\bf 91}, 053621 (2015).}

\bibitem{Bardyn2016}
C.-E. Bardyn, T. Karzig, G. Refael, and T. C. H. Liew, {\it Chiral Bogoliubov excitations in nonlinear bosonic systems}, Phys. Rev. B {\bf 93}, 020502 (2016).

\bibitem{Everitt2017}
P. J. Everitt et al., {\it Observation of a modulational instability in Bose-Einstein condensates}, Phys. Rev. A {\bf 96}, 041601(R) (2017).

\bibitem{Nguyen2017}
J. H. V. Nguyen, D. Luo, and R. G. Hulet, {\it Formation of matter-wave soliton trains by modulational instability}, Science {\bf 356}, 422 (2017).

\bibitem{MI_gauge}
K. Lelas, O. \v{C}elan, D. Prelogovi\'{c}, H. Buljan, and D. Juki\'{c}, {\it Modulation instability in the nonlinear Schr\"odinger equation with a synthetic magnetic field:  gauge matters}, Phys. Rev. A {\bf 103}, 013309 (2021). 


\bibitem{bardyn2014}
C.-E. Bardyn, S.D. Huber, and O. Zilberberg, {\it Measuring topological invariants in small photonic lattices}, New J. Phys. {\bf 16}, 123013 (2014).

\bibitem{aidelsburger2015}
M. Aidelsburger et al., {\it Measuring the Chern number of Hofstadter bands with ultracold bosonic atoms}, Nature Phys. {\bf 11}, 162 (2015).

\bibitem{wimmer2017}
M. Wimmer, H.~M. Price, I. Carusotto, and U. Peschel, {\it Experimental measurement of the Berry curvature from anomalous transport}, Nature Phys. {\bf 13}, 545 (2017).

\bibitem{tarnowski2019}
M. Tarnowski, F. N. \"Unal, N. Fl\"aschner, B. S. Rem, A. Eckardt, K. Sengstock, and C. Weitenberg, {\it Measuring topology from dynamics from obtaining the Chern number from a linking number}, Nature Commun. {\bf 10}, 1728 (2019).

\bibitem{Poshakinskiy2015}
A. V. Poshakinskiy, A. N. Poddubny, and M. Hafezi, {\it Phase spectroscopy of topological invariants in photonic crystals}, Phys. Rev. A {\bf 91}, 043830 (2015).

\bibitem{chong}
W. Hu, J.~C. Pillay, K. Wu, M. Pasek, P.~P. Shum, and Y.~D. Chong, {\it Measurement of a topological edge invariant in a microwave network}, Phys. Rev. X {\bf 5}, 011012 (2015).

\bibitem{mittal2016}
S. Mittal, S. Ganeshan, J. Fan, A. Vaezi, and M. Hafezi, {\it Measurement of topological invariants in a 2D photonic system}, Nature Photon. {\bf 10}, 180 (2016).


\bibitem{NDC}
R. W. Bomantara, W. Zhao, L. Zhou, and J. Gong, {\it Nonlinear Dirac cones}, Phys. Rev. B {\bf 96}, 121406(R) (2017).


\bibitem{Foesel2017}
T. F\"osel, V. Peano, and F. Marquardt, {\it L lines, C points and Chern numbers: understanding band structure topology using polarization fields}, New J. Phys. {\bf 19}, 115013 (2017).

\bibitem{leaky}
D. Leykam and D. A. Smirnova, {\it Probing bulk topological invariants using leaky photonic lattices}, Nature Phys. (2021). https://doi.org/10.1038/s41567-020-01144-5

\bibitem{Smerzi}
P. Buonsante, R. Franzosi, and A. Smerzi, {\it Phase transitions at high energy vindicate negative microcanonical temperature}, Phys. Rev. E {\bf 95}, 052135 (2017).

\bibitem{Flach}
T. Mithun, Y. Kati, C. Danieli, and S. Flach, {\it Weakly Nonergodic Dynamics in the Gross-Pitaevskii Lattice}, Phys. Rev. Lett. {\bf 120}, 184101 (2018).

\bibitem{Hafezi}
M. Hafezi, P. Adhikari, and J. M. Taylor, {\it Chemical potential for light by parametric coupling}, Phys. Rev. B {\bf 92}, 174305 (2015).

\bibitem{Wu2019}
F. O. Wu, A. U. Hassan, and D. N. Christodoulides, {\it Thermodynamic theory of highly multimoded nonlinear optical systems}, Nature Photon. {\bf 13}, 776 (2019).

\bibitem{FQH}
T. Neupert, L. Santos, C. Chamon, and C. Mudry, {\it Fractional Quantum Hall States at Zero Magnetic Field}, Phys. Rev. Lett. {\bf 106}, 236804 (2011).

\bibitem{TI_review}
\revision{M. Z. Hasan and C. L. Kane, {\it Colloquium: Topological insulators}, Rev. Mod. Phys. {\bf 82}, 3045 (2010).}

\bibitem{Berry_review}
\revision{D. Xiao, M.-C. Chang, and Q. Niu, {\it Berry phase effects on electronic properties}, Rev. Mod. Phys. {\bf 82}, 1959 (2010).}


\bibitem{Trager2006}
D. Tr\"ager et al., {\it Nonlinear Bloch modes in two-dimensional photonic lattices}, Opt. Exp. {\bf 14}, 1913 (2006).

\bibitem{NL_review}
Y. V. Kartashov, B. A. Malomed, and L. Torner, {\it Solitons in nonlinear lattices}, Rev. Mod. Phys. {\bf 83}, 247 (2011).

\bibitem{supp}
See Supplemental Material, which contains details of the linear stability analysis, calculation of the nonlinear Bloch wave spectrum, and simulations of the instability dynamics.

\bibitem{Shen}
\revision{S.-Q. Shen, \emph{{Topological Insulators. Dirac Equation in Condensed Matters}} (Springer, 2012).}

\bibitem{footnote2}
\revisiontwo{In Sec. 2 of Ref.~\cite{supp} we provide a justification of the perturbation modes maintaining a similar polarization for weak nonlinearities. In Sec. 3.2 of Ref.~\cite{supp} we plot examples of the perturbation modes’ wavevector-dependent spin and polarization in the trivial and non-trivial phases.
}

\bibitem{Hu2016}
Y. Hu, P. Zoller, and J. C. Budich, {\it Dynamical Buildup of a Quantized Hall Response from Nontopological States}, Phys. Rev. Lett. {\bf 117}, 126803 (2016).

\bibitem{Bardyn2013} 
C.-E. Bardyn, M. A. Baranov, C. V. Kraus, E. Rico, A. Imamoglu, P. Zoller, and S. Diehl, {\it Topology by dissipation}, New J. Phys. {\bf 15}, 085001 (2013).

\bibitem{Budich2015}
J. C. Budich and S. Diehl, {\it Topology of density matrices}, Phys. Rev. B {\bf 91}, 165140 (2015).

\bibitem{footnote1}
Alternatively, one could use a finite lattice superimposed with a sufficiently broad trapping potential.



\end{thebibliography}

\begin{thebibliography}{99}

\bibitem{sTI_review}
M. Z. Hasan and C. L. Kane, {\it Colloquium: Topological insulators}, Rev. Mod. Phys. {\bf 82}, 3045 (2010).

\bibitem{sBerry_review}
D. Xiao, M.-C. Chang, and Q. Niu, {\it Berry phase effects on electronic properties}, Rev. Mod. Phys. {\bf 82}, 1959 (2010).

\bibitem{sFoesel2017}
T. F\"osel, V. Peano, and F. Marquardt, {\it L lines, C points and Chern numbers: understanding band structure topology using polarization fields}, New J. Phys. {\bf 19}, 115013 (2017).

\end{thebibliography}
\end{document}